\documentclass{appolb}
\usepackage{graphicx}

\def \be {\begin{equation}}
\def \ee {\end{equation}}
\def \bea {\begin{eqnarray}}
\def \eea {\end{eqnarray}}
\def \nn {\nonumber}

\def \e {{\rm e}}

\def \n {\nu}

\def \ep{\epsilon}

\def \lab #1 {\label{#1}}

\def \endoc {\end{document}}

\def \p {\pi}
\def \m {\mu}

\def \as {{\alpha_s}}
\def \ra {\rightarrow}
\def \vep {\varepsilon}
\def \s {\sigma}

\ifx\pdfoutput\undefined
\input epsf

\else

\fi

\begin{document}

\title{Two Lectures on QCD at Short Distances: \\Jets and Factorization%
\thanks{YITP-SB-14-52.   Based on lectures presented at 54th Cracow School of Theoretical Physics.   This work was supported in part
by the National Science Foundation, award PHY-1316617.}%
}
\author{George Sterman
\address{C.N. Yang Institute for Theoretical Physics\\ Stony Brook University, Stony Brook NY 11794-3840 USA}
\\
%{Third Author of different affiliation
%}
%the Name(s) of other Author(s)
%\address{affiliation}
}
\maketitle
\begin{abstract}
This is a brief introduction to two of the central concepts in perturbative quantum chromodynamics, jets and factorization, which serve as windows into the short-distance behavior of quantum fields.
\end{abstract}
  
\section{Rare but Highly Structured Events}

In particle collisions at energies much above the mass of the proton,
certain exceptional final states include subsets of particles of momenta $\{q_j\}$ with anomalously small 
invariant mass: 
$( \sum_i q_i )^2 \ll ( \sum_i E_i )^2$, emitted at wide angles to the beam direction(s) and  not embedded among
other particles of similar energy.   Such sets of particle are  ``jets".
 Jets are a signature of large momentum transfer through local interactions, and
as such are direct evidence of processes taking place over distances
of the order of 1/(momentum transfer).     Proton collisions from all high energy colliders, and most recenty the Large Hadron Collider (LHC), provide striking examples of this phenomenon on a regular basis.

The first  lecture begins with a brief historical review of such particle jets, discusses them as a window to short distances, and attempts to explain why and when jets are seen.   It goes on to 
a discussion of the concept of infrared safety, its essential relation to energy flow, and to a description of a classical-quantum connection at the heart of jet phenomenology.
The second lecture describes how infrared safety is used in practice, and aims to communicate the flavor of the all-orders reasoning that underlies the proofs of infrared safety for jet cross sections in electron-positron annihilation and of their factorization in hadron-hadron scattering. 
\begin{figure}[htb]
\centerline{%
\includegraphics[width=8cm]{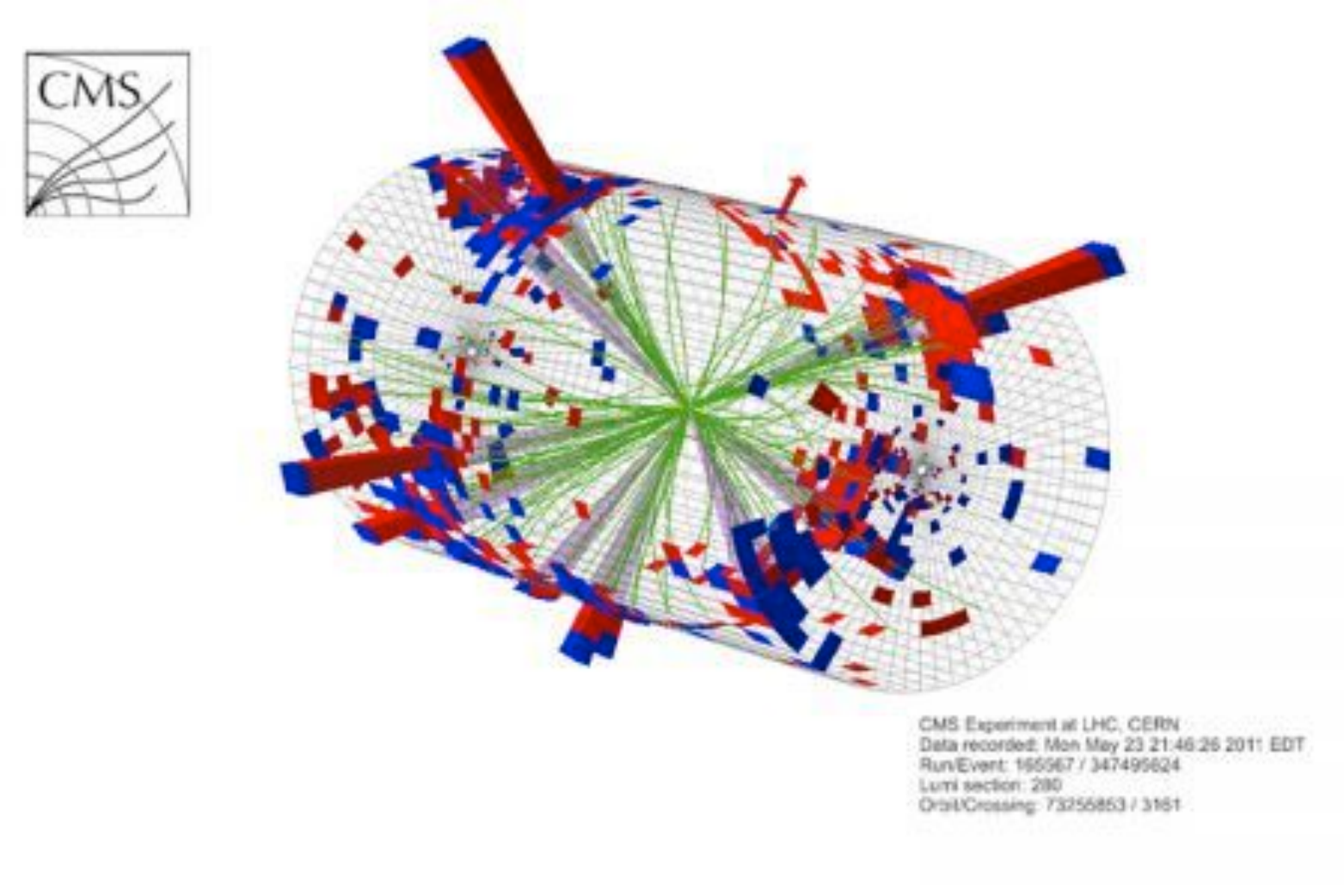}}%{cms_jet_set.pdf}}
\caption{An impressive multi-jet event recorded by the CMS Experiment at the LHC.}
\label{fig:cms-jet}
\end{figure}

The story is a mixture of history and perspective in the description of these rare but highly structured events.

\section{Particle  Jets, a Brief Biography}

  The prehistory of jets extends back to the 1950s,  in emulsion records of high energy cosmic rays.   Collisions of high energy projectiles with target nuclei resulted in highly directional collections of particles, which were dubbed jets.  In reports of one such experiment we read
 ``The average transverse momentum resulting from  our measurements is $p_T$=0.5 BeV/c for pions \dots Table 1 gives a
 summary of jet events observed to date \dots" \cite{Edwards:1957}.   We now interpret such jets as collections of fragments, following the directions of the projectile arriving from space in these experiments.   They were not quite the same as the wide-angle jets that will be the main subject here, but are related nonetheless, in ways we will see below.
 
 The observation of  ``wide angle"  or ``high $p_T$" jets, like the ones in Fig.\ \ref{fig:cms-jet} had to await the  era of high energy physics, beginning in the late 1960s. The door was opened by inclusive deep-inelastic scattering (DIS), and by the parton model which provided an explanation for its most striking feature, approximate scaling.    For the DIS cross section, scaling can be represented as
\bea
\sigma_{e-{\rm proton}}^{\rm inclusive}\left(Q,x\right) \ =\  \sigma^{\rm elastic}_{e-\rm parton} (Q) \times F_{\rm proton}(x)\, ,
\label{eq:scaling}
\eea
in which the {\it inclusive} electron-proton cross section at spacelike momentum transfer $q^\mu$, with $-q^2=Q^2$, is given by the {\it elastic} scattering cross section of an electron on a point-like charged particle times a function $F_{\rm proton}$ of  a dimensionless variable $x=Q^2/2p\cdot q$.  Simple kinematic reasoning suggests that the ``scaling variable" $x$  has the kinematic interpretation of the fraction of the proton's momentum carried by the elastically-scattered point particle, as seen in the rest frame of the incident electron.   These momentum fractions were insightfully assigned to hypothetical partons, the probabilities of whose fractional momenta are described by $F_{\rm proton}(x)$, the ``parton distributions."  \cite{feynpm,bjorken69}.
 
This was a beautiful but puzzling explanation, given that the leading candidates for charged partons were quarks, which by then were thought to be confined.  But if they were confined, how could they appear to scatter elastically?   It wasn't too long, however, before this paradoxical combination of properties was seen to emerge naturally from quantum field theory through the asymptotic freedom of quantum chromodynamics (QCD).  Subsequently, the charged partons were unambiguously identified with quarks, in part from the spin-dependence of the electron-parton elastic scattering in Eq.\ (\ref{eq:scaling}) \cite{asyfree}.  The question then quite naturally arose, what happens to partons in the final state?   Wouldn't confinement, the reverse side of the coin of asymptotic freedom, still forbid a dynamical expression of quarks and color?   The detectors and energies available for these pioneering experiments did not allow immediate answers to these questions.

\begin{figure}[htb]
\centerline{%
\includegraphics[width=5cm]{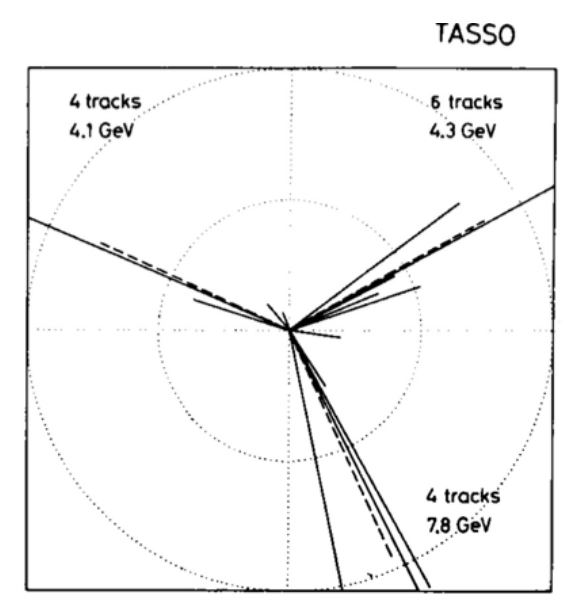}}
\caption{A three-jet event, including a gluon jet, as seen by a Petra experiment \cite{Wu:1984ik}.}
\label{fig:gluon-jet}
\end{figure}

The answer, however, was not long in coming from SLAC, in 1975.    Confinement did not prevent a direct signature of quark-pair production in electron-positron annihilation to hadrons.    The angular distribution of energy flow in such events follows the Born expression for the creation of spin-1/2 pairs of quarks and antiquarks \cite{han75}.    In QCD, of course, there are gluons as well as quarks, and a few years later there were hints of three gluon structure  in Upsilon decay \cite{Berger:1978rr}, and then unequivocal gluon jets were found at the higher-energy electron-positron machine, Petra \cite{Wu:1984ik}, very much as described in Refs.\ \cite{Ellis:1976uc,Ellis:1978wp}, and as illustrated in Fig.\ \ref{fig:gluon-jet}.   These observations confirmed color as a dynamical variable.

\begin{figure}[htb]
\centerline{%
\includegraphics[width=8cm]{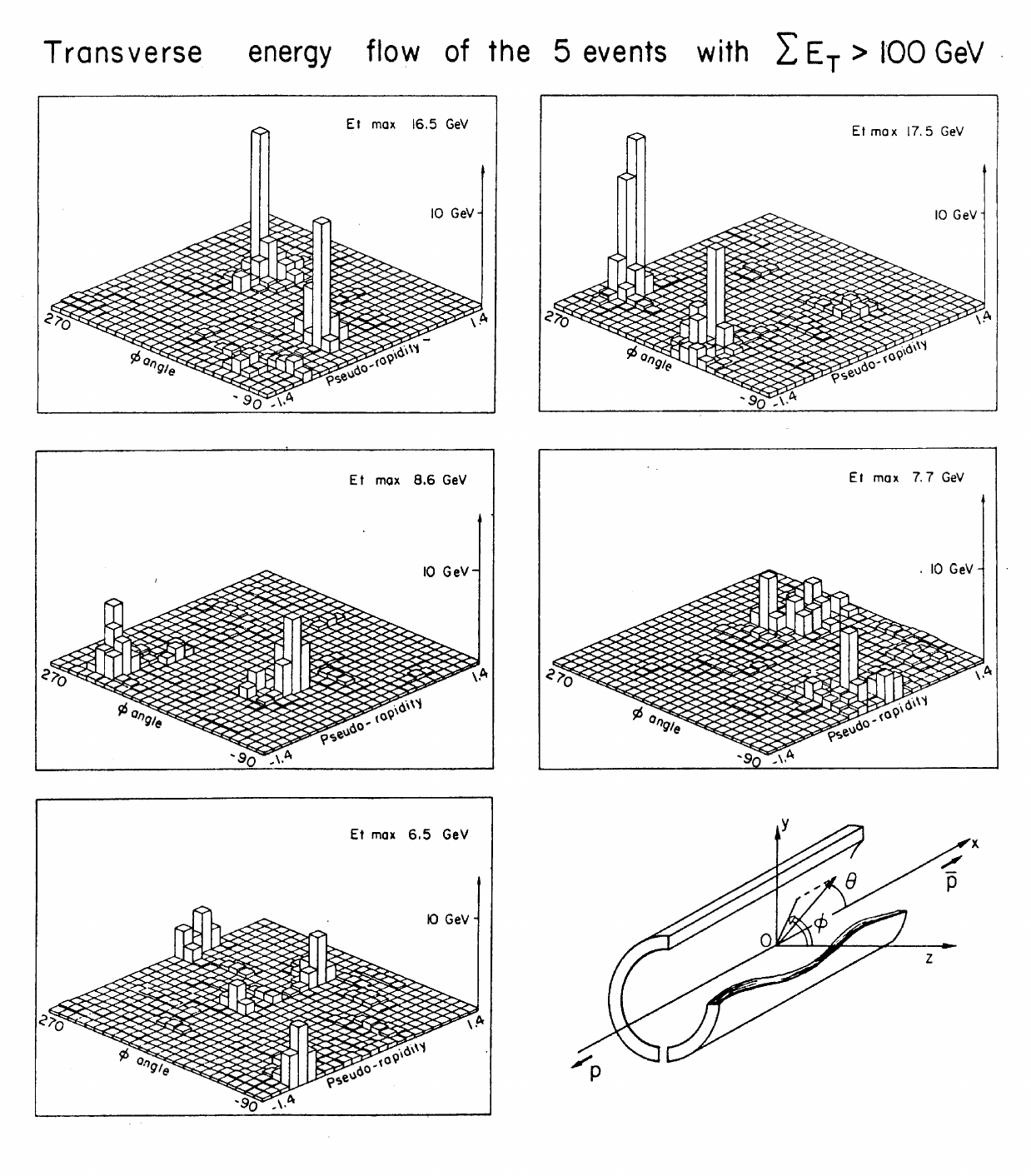}}
\caption{UA1 ``Lego plots", representing the flow of energy in a rolled-out representation of the detector surface \cite{Arnison:1983rn}.}
\label{fig:ua1-rev}
\end{figure}

 The next generation of machines allowed hadron-hadron scattering, and by implication the scattering of quarks and gluons.
In the early  1980s, there was strong evidence for  scattered parton jets at the ISR \cite{Della Negra:1977sk} and Fermilab \cite{Corcoran:1979xt}.
Overall, however, it was an unsettled period until the  large angular coverage of detectors at the SPS made possible ``lego plots" in terms of energy flow, and 
provided the clear observation of high-$p_T$ jet pairs that represent the scattering of partons by partons, mediated by the strong force itself, as in Fig.\ \ref{fig:ua1-rev} \cite{Arnison:1983rn}.   Such jets are naturally ``rare" because the high momentum transfer scattering of partons in QCD is rare (but calculable), decreasing as a power of the energy transfer on dimensional grounds.   

\begin{figure}[hb]
\centerline{%
\includegraphics[width=6cm]{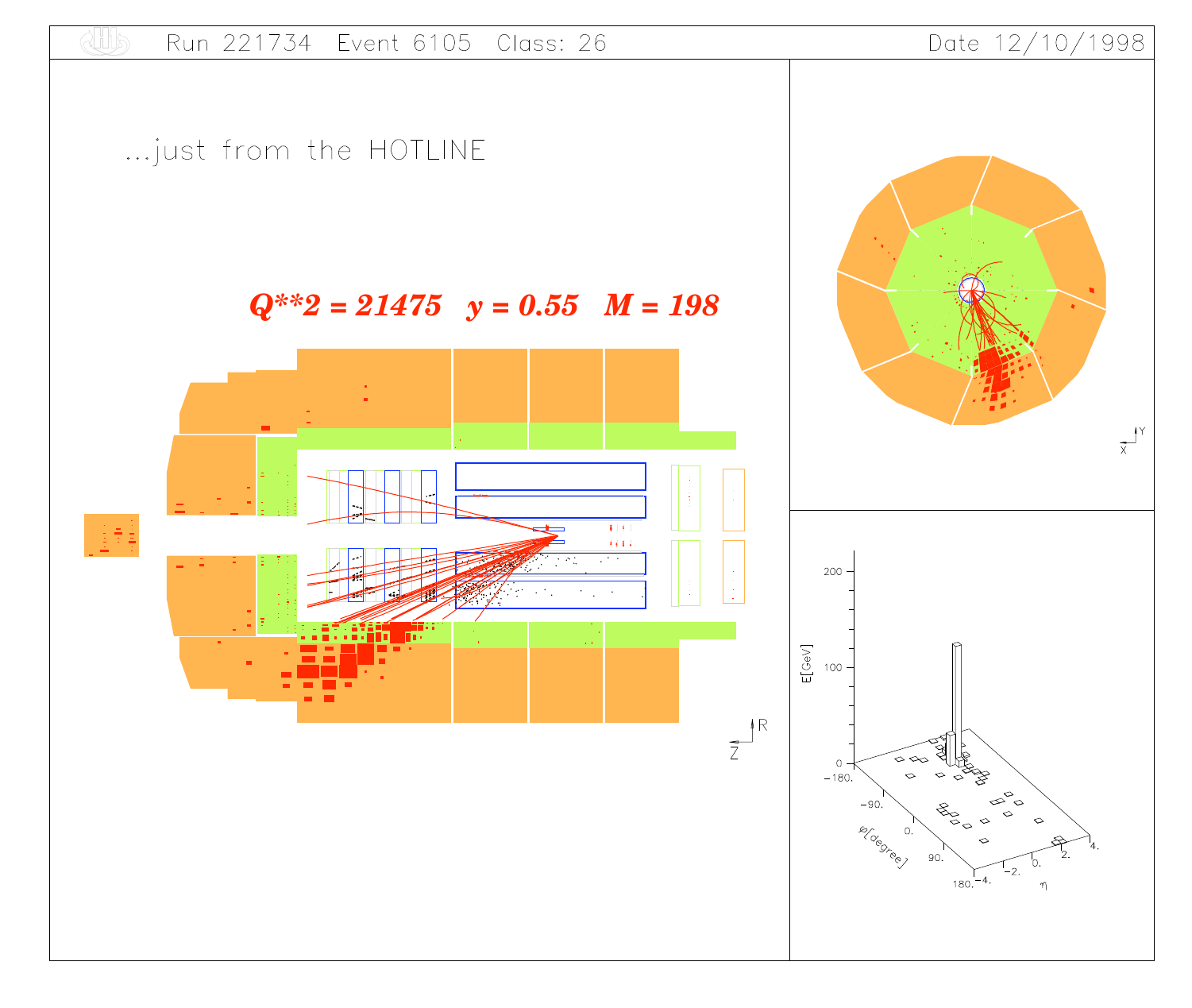}}
\caption{A DIS event seen by the H1 experiment at HERA in side view and Lego plot.}
\label{fig:dis-jet}
\end{figure}

\begin{figure}[hb]
\centerline{%
\includegraphics[width=6cm]{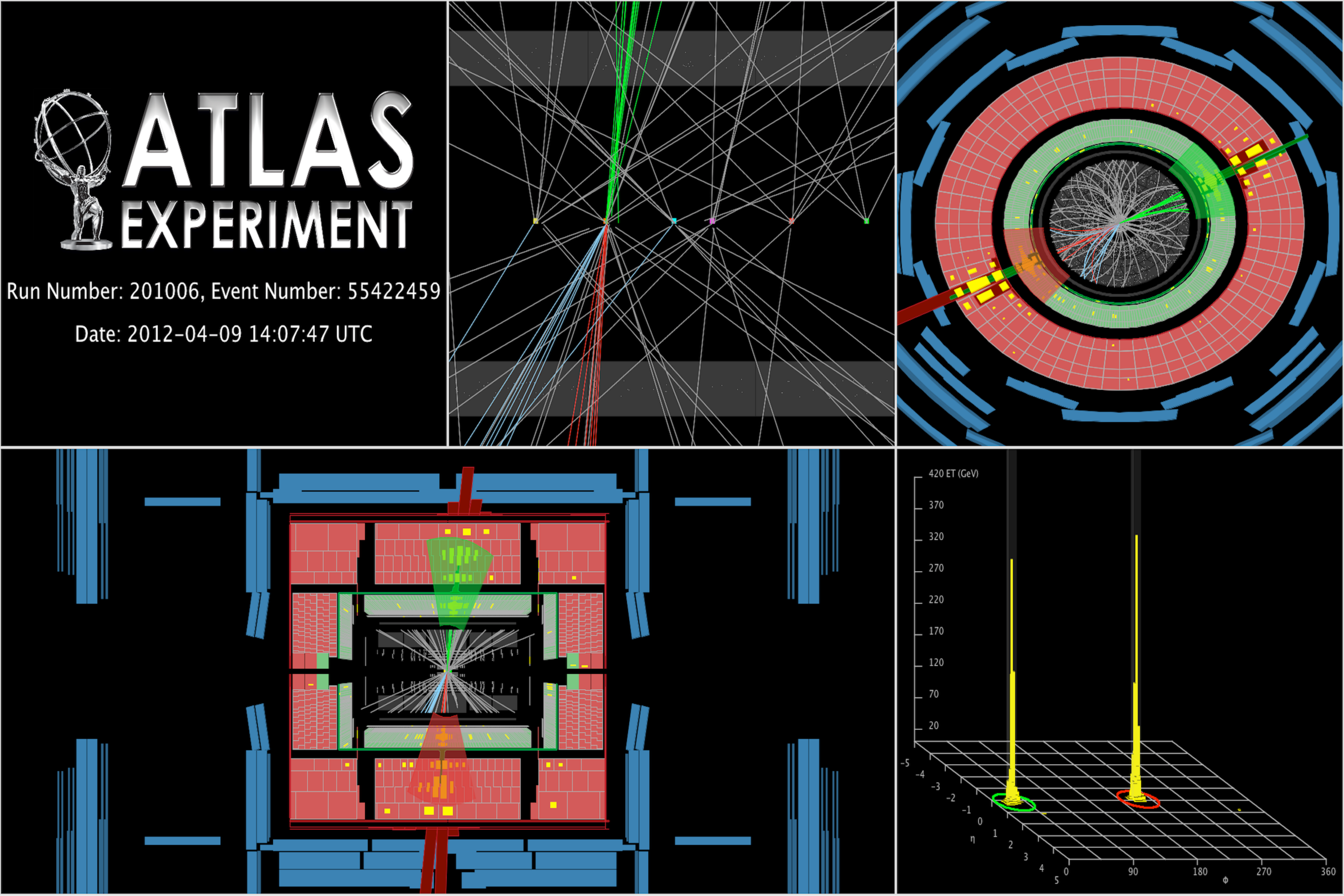}}%{very-high-pt}}
\caption{A Terascale jet seen by the Atlas Experiment.}
\label{fig:atlas-jet}
\end{figure}

\begin{figure}[htb]
\centerline{%
\includegraphics[width=10cm]{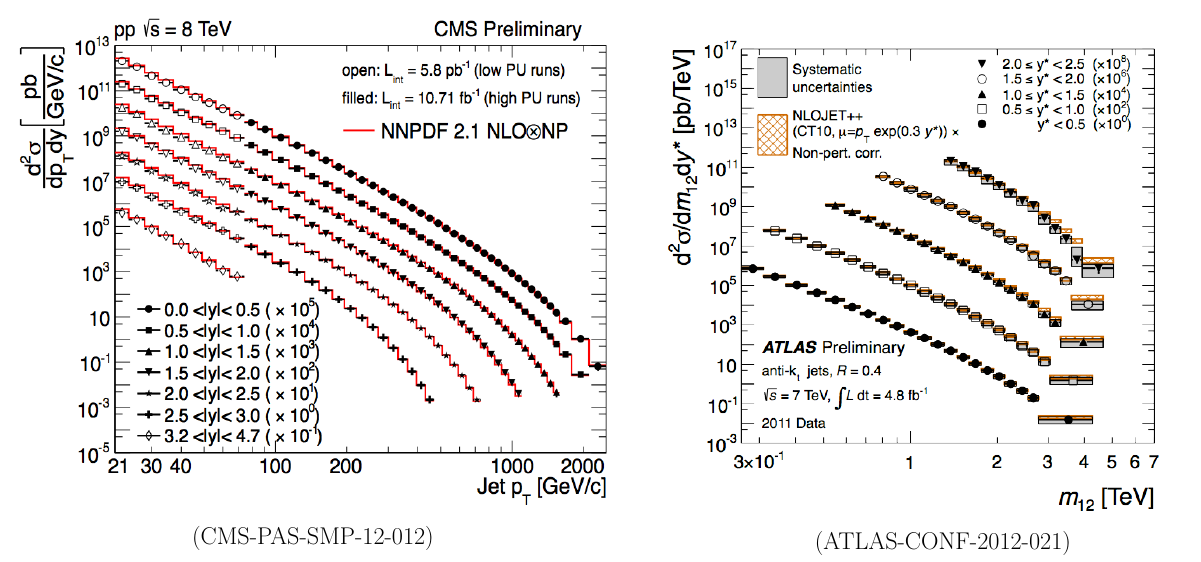}} 
\caption{Single-jet inclusive and jet pair mass distributions measured by the CMS and Atlas Experiments over many orders of magnitude.}
\label{fig:jet-sigma}
\end{figure}

\begin{figure}[htb]
\centerline{%
\includegraphics[width=10cm]{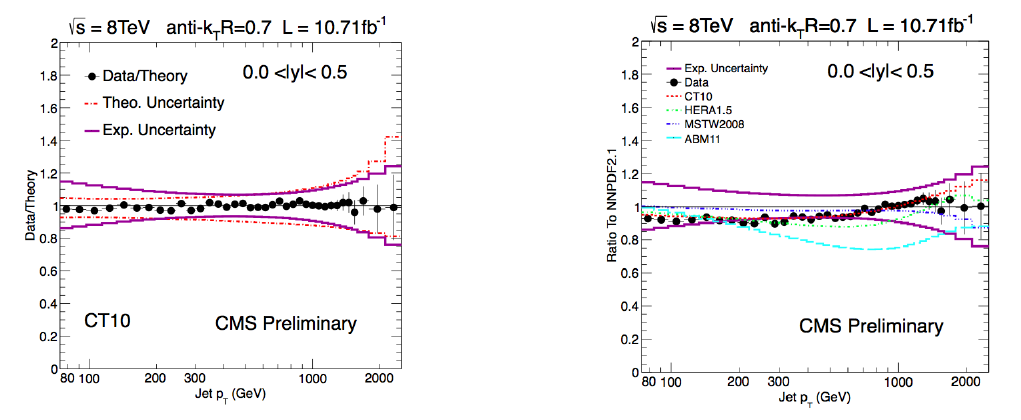}}\
\caption{Single-jet inclusive cross sections measured by the CMS Experiment, in ratios of theory to experiment.}
\label{fig:jet-theory-to-exp}
\end{figure}

In fast-forward, we move to the 1990's, and the machines that truly established the Standard Model:   HERA, the Tevatron Run I, and LEP I and II.   Here,
jet cross sections were seen and measured over multiple orders of magnitude.   At HERA, the scattered quark of Eq.\ (\ref{eq:scaling}) makes a clear appearance, as illustrated by Fig.\ \ref{fig:dis-jet} from the H1 experiment.   Finally, this century has seen a new era for jets at the limits of the Standard Model, ushered in by Tevatron Run II, and  then the LHC 7 $\rightarrow$ 8, heading towards 14 TeV, with jets whose energies reach 1 TeV and beyond, as in Fig.\ \ref{fig:atlas-jet}.   Basic uncertainty relations suggest that such events emerge from processes at distance scale $\delta x \sim \frac{\hbar c}{1 \ {\rm TeV}} \sim 2\times 10^{-19}$ meters.  They are observed by detector elements up to about 10 meters away, thus bridging 20 orders of magnitude in a single experiment.   
Over time, theory has kept up with these developments, as illustrated by Figs.\ \ref{fig:jet-sigma} and \ref{fig:jet-theory-to-exp}, which show remarkable consistency of theory to data in both shape and normalization, over more than ten orders of magnitude in the cross section.

\section{Jets as a Window to Short Distances}

\begin{figure}[htb]
\centerline{%
\includegraphics[width=8cm]{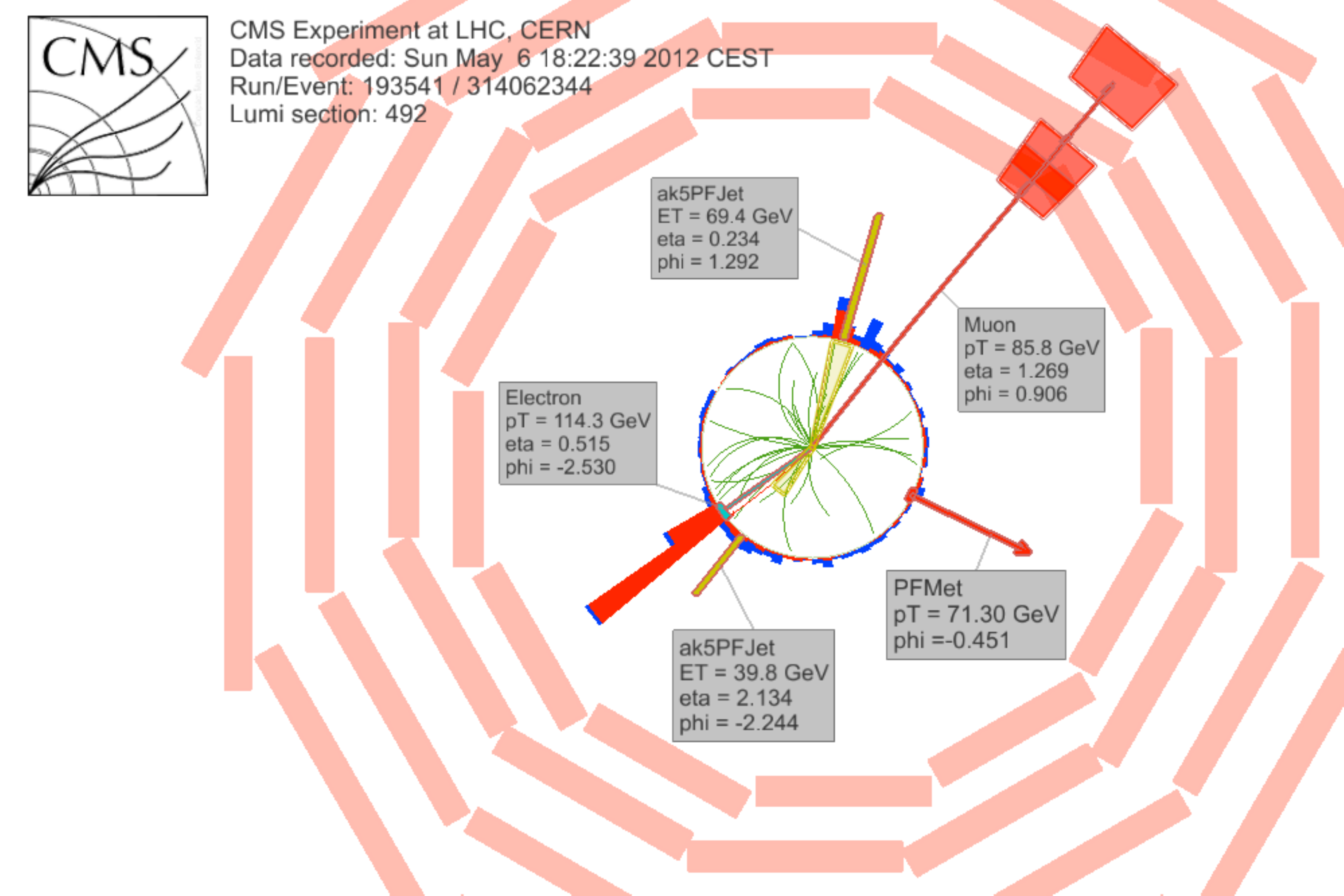}}%{top-fig_atlas-2010-063-fig_09}}
\caption{A candidate top quark pair production event.}
\label{fig:top-event}
\end{figure}

The jets seen to date are, as the data shown above demonstrate, primarily QCD phenomena, and this very fact provides strong tests of the Standard Model.   For example, limits on the compositeness of quarks are set by observing ``Rutherford-like" scattering of quarks by gluon exchange.   Data show that the rapidity distribution tracks the predictions of a process whose lowest order is the exchange of a single gluon between two point sources.  Similarly, by reconstructing sets of jets, often in combination with observed leptons, it is possible to observe, and sometimes discover, heavy particles, like the top quark, Fig.\ \ref{fig:top-event}.

Most of all at the LHC, jets are studied as backgrounds to and signals for new physics.  For example, a gluino, the spin-1/2 superpartner of the gluon, would decay into a squark, the superpartner of a quark, by radiating an antiquark, which appears as a jet in the final state.   Then the squark decays to a stable, weakly-interacting supersymmetric particle by radiating an antiquark, which appears in the final state as another jet.   Since in most supersymmetric models, gluinos, should they exist, can be produced only in pairs, the resulting final state has at least four jets and two sources of missing energy (the lightest supersymmetric particle).  Such an event is complicated, but by no means uniquely produced by supersymmetry.  The Standard Model produces just such events, and distinguishing between this background and that signal requires a level of precision in the calculation of each.

Quite generally, jets appear whenever a short-lived state, whether QCD, electroweak, or new physics,  decays to strongly-interacting matter.    In the following we discuss how this comes about.

\section{Why and When Are There Jets?}

The calculation of jet production, and indeed any cross section in QCD, requires learning to tease predictions out of a quantum field theory that acts differently on different length, and correspondingly momentum, scales.   One solution to this problem is ``factorization" \cite{cssrv}.    For factorizable processes, we can separate the calculable  short-distance scattering of quarks and gluons from the long-distance binding of hadrons.  Jet cross sections are only one among a larger class of observables for which this is possible.    A sibling area, for example, is the study of elastic scattering, even rarer events in which jets are replaced by single hadrons, (like $p+p\rightarrow p+p$)  \cite{qcdform}.   Factorization techniques for these and other processes are still being developed, often in combination of ideas from effective field theory \cite{scet}.  

 A factorized jet cross section is a direct generalization of the expression for deep-inelastic scattering, Eq.\ (\ref{eq:scaling}) \cite{cssrv,Bodwin:1984hc}
\bea
{d\sigma(A+B\rightarrow \{p_{c_i}\}) }
&=&  \int dx_adx_b \ {{f}_{a/A}(x_a,\m_F)\, {f}_{b/B}(x_b,\mu_F)}\ 
 \nonumber\\
 &\ & \hspace{-25mm}
 \times \ \ {C\left(x_ap_A,x_bp_B,\frac{Q}{\mu_F},p_{c_i}\right)_{ab\rightarrow c_1 \dots c_{N_{\rm jets}+X}}}
 d\ \left[ {\prod_{i=1}^{N_{\rm jets}}\ J_{c_i}(p_{c_i},\m_F)} \right]\, .
 \nn\\
  \label{eq:jet-fact}
\eea
The elements of this expression are parton distributions, $f_{a/A}(x_a,\m_F)$, short distance ``coefficients" $C\left(x_ap_A,x_bp_B,\frac{Q}{\mu_F},p_{c_i}\right)_{ab\rightarrow c_1 \dots c_{N_{\rm jets}+X}}$, and final-state functions of the jet momenta, $J_{c_i}(p_{c_i},\m_F)$.    Together, they  tell a story, in which partons $a$ and $b$, whose probability distributions are given by the $f$s, collide through a set of calculable quantum mechanical processes described by the $C$s, resulting in particles that evolve independently into the final state to produce jets, with probability distributions in their own momenta given by the $J$s.  The jets, of course, are made up of individual particles, and  very closely related to these jet cross sections are factorized single-particle inclusive cross sections,
\bea
{d\sigma(A+B\rightarrow H(p)) }/{d^3p}
&=&  \int dx_adx_b \  {{f}_{a/A}(x_a,\m_F)\, {f}_{b/B}(x_b,\mu_F)}\ 
 \ \nn \\
 &\ & \hspace{5mm}
 \times \ \ \int dz\ {C\left (x_ap_A,x_bp_B,\frac{p}{z\mu_F}\right)_{ab\rightarrow c(p/z)}}
\nonumber\\
 &\ & \hspace{5mm}
  \times\ \  D_{H/c}(z)\, ,
  \label{eq:1pi}
\eea
where the ``fragmentation function" $D_{H/c}(z)$ is the probability distribution for hadrons $H$ to appear in a ``$c$-jet", with $z$ the fractional momentum of the jet carried by hadron $H$.   

To interpret these fundamental results is to address the questions: Why are there jets in quantum field theory?  Why is the dynamics of jets factorized, or one might say, ``autonomous"?   In the following, we will try to answer these questions in turn,  partly with quantum mechanical, partly with classical reasoning.  

\subsection{The Why of Jets} 

Let's go back to some of the very basics of quantum field theory in its perturbative description.    All perturbation theory arises from a Schr\"odinger equation that controls mixing of (free particle) states, when we are given a free Hamiltonian $H^{(0)}$, and a perturbing ``potential", $V$, representing a set of operators that mix the free states,
\bea
\hspace{-30mm}
i\hbar\, {\partial \over \partial t}|\psi(t)>= \left( H^{(0)}+V\right)|\psi(t)>\, .
\label{eq:SEqn}
\eea
We solve this equation with free-state ``in"  boundary conditions,  
\bea
|\psi(t=-\infty)>=|m_0> =  | p_1^{\rm IN}, p_2^{\rm IN}\rangle \, .
\label{eq:in-states}
\eea
Here we use the notation
$V_{ji}=\langle m_j|V|m_i\rangle$, which specifies the vertices in the diagrammatic picture of our theory.  Theories differ by their lists of particles that constitute the free states $|m\rangle$, and by the lists of their (hermitian) $V$s.

Solutions to the Schr\"odinger  equation (\ref{eq:SEqn}) are sums of  ordered time integrals, and taking the matrix element of such a solution with an ``out" state $\langle m_n|$, we find the S matrix,
\bea
\langle m_n| m_0\rangle &=& \sum_{\tau\, {\rm orders}} 
\int_{-\infty}^\infty d\tau_n \dots \int_{-\infty}^{\tau_2} d\tau_1
\nonumber\\
\nn\\
&\ & \hspace{5mm} \times \prod_{{\rm loops}\, i} \int {d^3\ell_i\over (2\pi)^3}\ \prod_{{\rm lines}\,  j}{1\over 2E_j}
 \times \ \prod_{{\rm vertices}\ a} i {V_{a \rightarrow a+1}}
\nonumber\\
\nn\\
&\ & \hspace{5mm} \times 
\exp {\left[\;  i\ \sum_{{\rm states}\, m}
\left(\, {\sum_{j\, {\rm in}\, m} E(\vec p_j)}\right) (\tau_m-\tau_{m-1})\right]}\, .
\label{eq:topt}
\eea
This is sometimes referred to as ``old-fashioned perturbation theory".  The $E(\vec p_i)=\sqrt{\vec p_i\, {}^2+m_i^2}$ are the energies of the particles in each intermediate state.  Because spatial momentum (but not energy) is conserved at each vertex, the integrals over loop momenta $\ell_i$ are equivalent to sums over free field intermediate states.
 
 The time integrals in Eq.\ (\ref{eq:topt}) extend to infinity, but in principle phase oscillations can damp them and answers can be finite.  Long-time, ``infrared" divergences, usually logarithmic, however, can come about when phases vanish over large stretches of the $\tau_i$ integrals, which can then grow at large times.
We would like to know when this happens.    

Some general insights can be found by looking at the phase on the right-hand side of (\ref{eq:topt}), which is given by
\bea 
\exp {\left[\;  i\ \sum_{{\rm states}\, m}
\left(\, \sum_{j\, {\rm in}\, m} E(\vec p_j)\right) (\tau_m-\tau_{m-1})\right]}
&=&
\nonumber\\
&\ & \hspace{-60mm} 
\exp {\left[\;  i\ \sum_{{\rm vertices}\, m}
\left(\sum_{j\, {\rm in}\, m} E(\vec p_j)\ -\ \sum_{j\, {\rm in}\, m-1} E(\vec p_j)\right)\ \tau_m\, \right]}\, .
\label{eq:phase}
\eea
Divergences for $\tau_i\rightarrow\infty$ require two things, which we can identify from the right- and left-hand sides of this expression.   From the right, we see that the vanishing of the phase requires degenerate states,
\bea
\sum_{j\, {\rm in}\, m} E(\vec p_j)= \sum_{j\, {\rm in}\, m-1} E(\vec p_j)\, .
\label{eq:degenerate}
\eea
This, however, is at most a necessary condition for divergence.   The phase is a fairly complicated function of the spatial momenta of particles in all the virtual states.  Divergence requires not only vanishing phase but also stationary phase, so that small variations in these momenta do not lead to large oscillations.  From the left of (\ref{eq:phase})  we derive a very useful condition for  stationary phase,
\bea
{{\partial \over \partial \ell_i{}_\mu} \; \left[\, {\rm phase}\, \right] 
=
 \sum_{{\rm states}\, m}\ \sum_{j\, {\rm in}\, m} \, (\pm \beta_j^\mu)(\tau_{m+1}-\tau_m)
 =0}\, ,
\label{eq:stationary}
\eea
where the $\beta_j$s are normal 4-velocities:
\bea
{\beta_j=\pm \partial E_j/\partial\ell_i}\, .
\label{eq:velocities}
\eea
Now the quantity $\beta^\mu \Delta \tau =x^\mu$ is the classical translation of an on-shell particle with velocity $\beta$ over time $\Delta \tau$.   Because Eq.\ (\ref{eq:stationary}) must hold for every loop whose vertices go far into the future, all vertices between every one of these loops must be connected with free, classical propagation between fixed points in space representing the vertices as $\tau \rightarrow \infty$. That is, the lines that give long-time behavior must describe a {\it physical process}.   This is actually very restictive, but easy to satisfy if all the $\beta_j$'s are equal in a given loop \cite{cssrv,colnor,ste78,Erdogan:2013bga}.  
 
 From these relatively simple considerations, we can draw a strong conclusion.     If  fast partons emerge from the same point in space-time,
 {{\it they can only rescatter with collinear partons}.}   But such an ensemble describes a jet of particles.    In essence, this is the answer to ``when are there jets?"   Jets require strong localization of the emission of a set of partons that propagate essentially on their mass shells to the far future.   This, of course, is natural when particles are produced at short distances in high energy collisions, and also in the decay of heavy to light particles.  On the other hand, this condition is very restrictive, and most sets of degenerate states cannot lead to divergent amplitudes.
  
\begin{figure}[htb]
\centerline{%
\includegraphics[width=6cm]{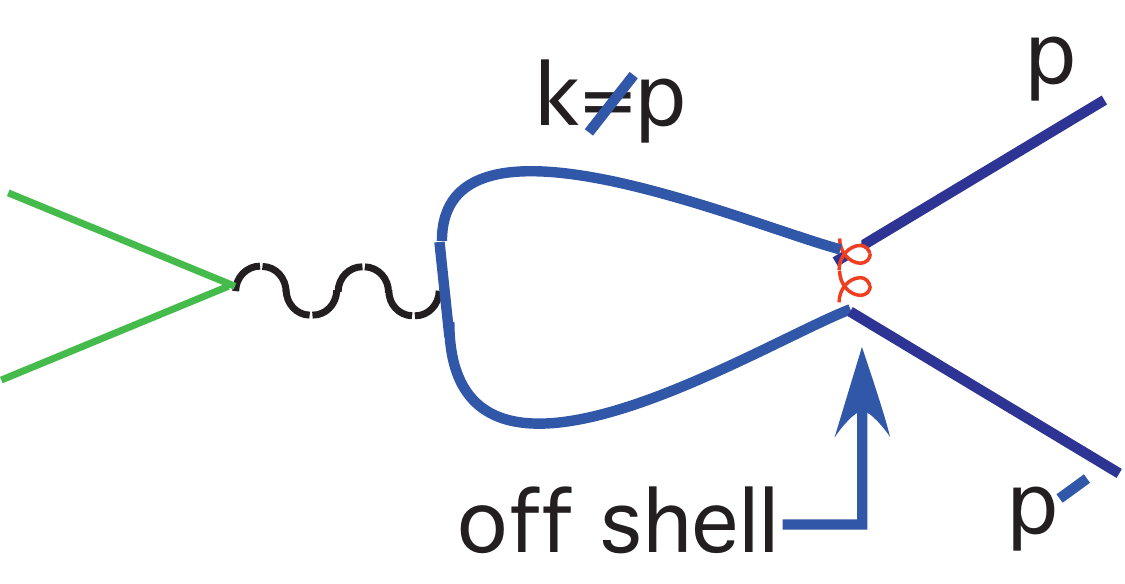}}
\caption{Degenerate states that do not lead to infrared divergences.}
\label{fig:no-pinch}
\end{figure}

 Let's illustrate the role of classical propagation.  An example of a set of degenerate states that cannot give long-time divergences is shown in Fig.\ \ref{fig:no-pinch}.   The pair of lines labelled $k \ne p$ can be degenerate with the final state $|p,p'\rangle$, so that the phase associated with this state vanishes, but it cannot give stationery phase unless the particles in the virtual pair are actually parallel to the external lines.   In terms of our discussion above, this is because the two virtual particles originate at the same point, and then, once they move for any finite time in different directions, they can never meet to scatter with the finite momentum transfer necessary to redirect them into the $p$ and $p'$ directions.   Thus, the effects of these states are limited to short times, of the order of the total energy of the pair, and their contributions to the amplitude are finite.

\begin{figure}[htb]
\centerline{%
\includegraphics[width=6cm]{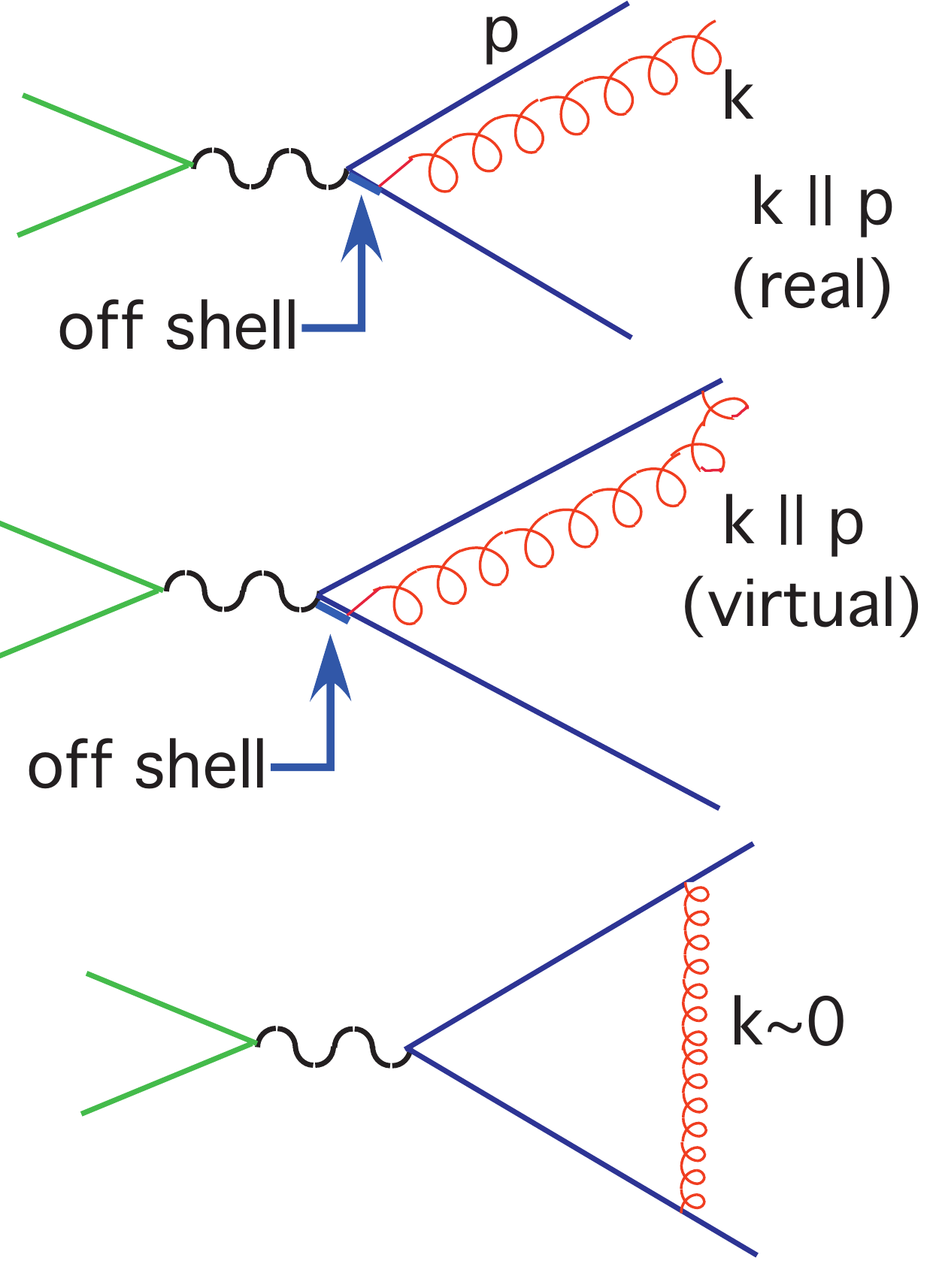}}
\caption{Degenerate states that can lead to infrared divergences.}
\label{fig:pinch}
\end{figure}

In contrast, for particles emerging from a local scattering as in Fig.\ \ref{fig:pinch}, (only) collinear (or soft) lines can give long-time behavior.   In the collinear case, two lines travel parallel to each other at the speed of light, and can interact locally at any time in the future.   The term ``soft" refers to the exchange of lines whose energy vanishes, and which therefore do not contribute to the phase in Eq.\ (\ref{eq:topt}).    In the space-time picture, these lines have infinite wavelength, and so can attach at any point in the physical picture.

These simple considerations generalize to any order, and to any field theory, and we conclude that the momentum configurations associated with jets are precisely those whose contributions to quantum mechanical amplitudes are large, and can even diverge.    We now turn to the independence of jet evolution and structure from the remainder of the event.

\subsection{The autonomy of jets}

In the factorized equations (\ref{eq:jet-fact}) and (\ref{eq:1pi}), the dependence of the cross sections on jet momenta, for example their invariant masses $p_c^2$  and $z$-dependence of their fragmentation functions,  appear in independent factors, which depend only on the nature of the partons that initiate the jets.   We will try to justify of this ``autonomy".  The first thing to point out is that  autonomy is by no means obvious.  What about the effect of long-range forces associated with massless gluons?  They might only transfer small momenta, but such effects might still modify jet substructure, and correlate it with the substructure of other jets.    We will give an intuitive argument for why this doesn't happen at high enough relative momenta between the jets, based on a simple classical analogy.    This will be followed by a sketch of more technical arguments later in the lectures.

 Let's think of the classical electromagnetic fields seen by a fast-moving particle receding from a charged particle.   A classical picture isn't so far-fetched, because the correspondence principle is the key to infrared radiation.   An accelerated charge must produce classical radiation, and an essentially infinite number of soft photons or  gluons is required to make up a classical field.

 In any case, we imagine that two particles are produced at the same place and time, which we choose as the origin in their respective rest frames, with coordinates $x^\mu$ and $x'{}^\mu$.   We are interested in the fields seen by the particle with the $x'$ frame as it recedes from the charged particle with  relative velocity $\beta$, which is near the speed of light, in the $x_3'$ direction.   All we need is the expression for the $x_3$ coordinate of the receding particle in the charged particle's rest frame, as a function of the time in its own frame,
 \bea
 x_3 \ =\  \sqrt{\frac{1}{1-\beta^2}}\; (x'_3\ +\ \beta t')   \ =\ \gamma\beta t'\, ,
\label{eq:x-3}
\eea
where we use that $\vec x{}\, '=0$ for the receding particle siting at the origin in its own frame.   In these terms, the 3-component of the electric field of the particle with charge $q$ as seen by the receding particle in its own coordinates is 
\bea
E_3'(x') 
\ =\  \frac{qx_3 }{ (x'_3\, {}^2)^{3/2} } \ \sim\  \frac{1}{\gamma^2}\, \frac{q}{(\beta t')^2} \, ,
\label{eq:field}
\eea
which decays not only as $1/t'\, {}^2$, but as an overall factor of $1/\gamma^2$, vanishing rapidly as $\beta\rightarrow 1$ unless $t'$ is very small.   As $\beta$ approaches the speed of light, then, the charge of one particle can exert an appreciable force on the other particle over only a very short time, much shorter than the time it would take for the outgoing particle (our proxy for a jet) to evolve into its eventual final state.   

From these considerations, it might seem that factorization should be easy to prove.   It's not quite this simple, however, if only because in perturbation theory the electric field is a derived quantity.   From a calculational point of view, the primary quantities are the gauge fields, $A^\mu$, whose Lorentz transformations are quite different from those of the field strengths. The vector potential, ${A}^\mu$ actually turns out to be uncontracted by the Lorentz transformation, but is mostly a total derivative as seen in the $x'$ frame,
\bea
\hspace{-10mm}
A'\, {}^\mu = q \frac{\partial}{\partial x'_\mu}\ \ln \left( \beta t'\right)
+
{\cal O}( 1 - \beta) \, .
\label{eq:gauge-field}
\eea
 Thus, the ``large" part of $A^\mu$  must be removed by a gauge transformation, something which requires special considerations in perturbation theory.   Once the total derivative in the gauge field is brought under control, however, the residual ``drag" forces are corrections of the general size 
\bea
1- \beta\ \sim\ \frac{1}{2}\; \left [\sqrt{1-\beta^2}\right ]^2\ \sim\ \frac{m^2}{2E^2}\, ,
\label{eq:h-t}
\eea
with $m$ the particle mass.  This gives a sense of the size of power suppressed corrections to ``jet autonomy", an estimate which is consistet with factorization, and which vanishes rapidly at large momentum transfer.

\begin{figure}[htb]
\centerline{%
\includegraphics[width=8cm]{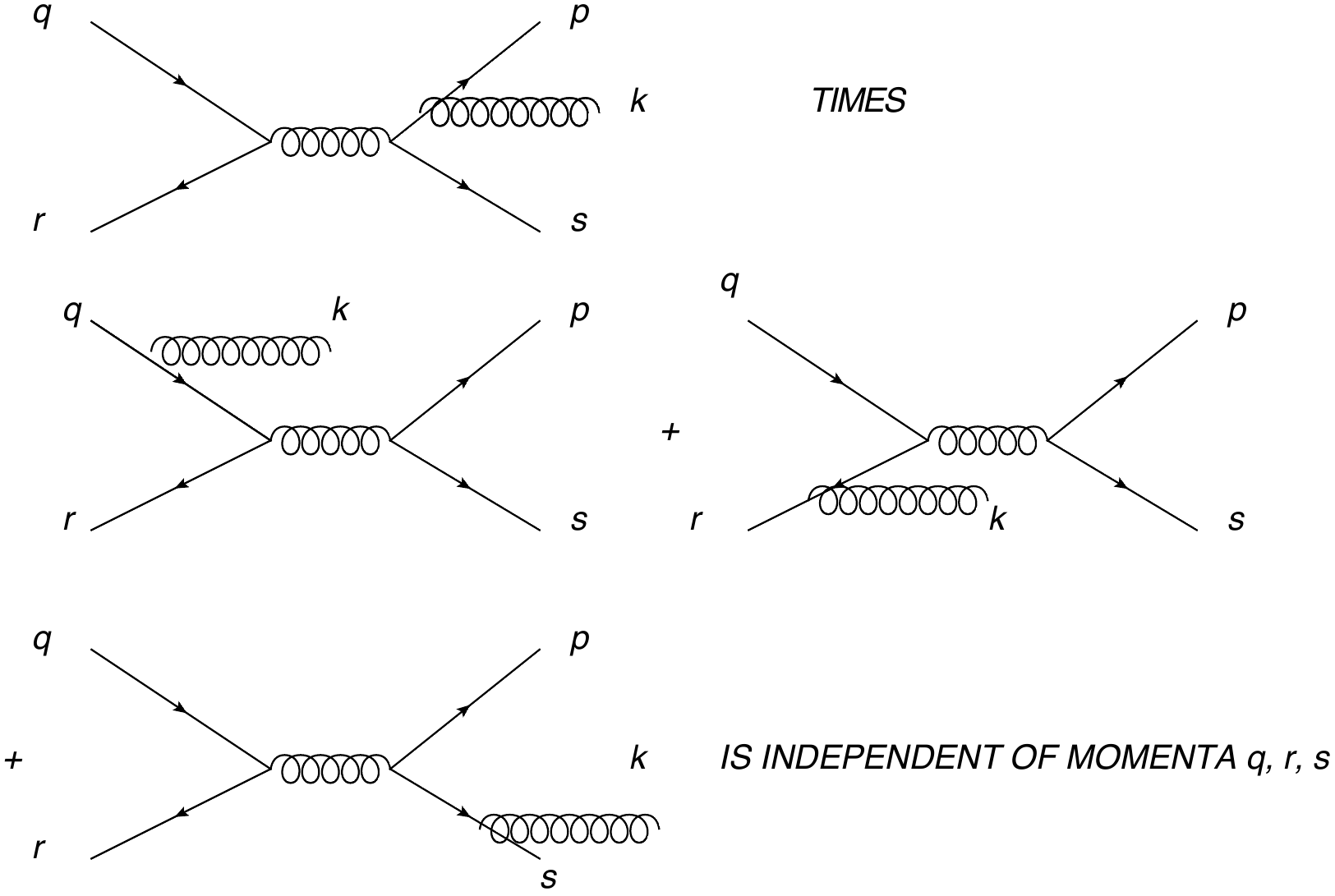}}
\caption{Interference contributions to the emission of a gluon ($k$) in a scattering process.  In the limit that $k$ becomes parallel to $p$, this contribution factorizes.}
\label{fig:interference}
\end{figure}

These general considerations apply as well to QCD.   We can consider, for example, the emission of a gluon of momentum $k$ in a scattering process $p+q\rightarrow r+s$, as illustrated in Fig.\ \ref{fig:interference}, which shows the interference between the emission of a gluon from line $p$ (top of the figure) with emission from the remaining lines.  In general, this contribution is complicated, but when $k$ is emitted nearly collinear to $p$, the sum of the three diagrams simplifies, and the entire contribution depends only on $p$ and $k$, and is naturally thought of as part of the jet associated with particle $p$ in a factorized cross section like Eq.\ (\ref{eq:jet-fact}).   The cross section is also large in this limit, consistent with the considerations of the previous subsection.   
 This simplification depends on gauge invariance, which is the key to the factorization of jets in QCD.   The jet only knows the rest of the world as a source of 
unphysically-polarized gluons.   

\subsection{Infrared Safety, Energy Flow and a Classical-Quantum Connection}

 We can now summarize:  we have found that amplitudes for the production of parallel-moving particles are greatly enhanced.   Each such set is a jet.   Massless particles can emit other massless particles, so the number of partons in any jet is indefinite and generally speaking amplitudes are ill-defined.    For massive lines, once $m \ll E$, amplitudes are enhanced, if not divergent.

We are thus faced with the problem of how to get quantitative predictions from divergent amplitudes.   In QCD, $t\rightarrow \infty$ surely invalidates perturbation theory because of nonperturbative effects, including confinement.  We want to use scattered partons,  but we can't follow them to infinity, and that's the problem.   Here's another analogy, this time from quantum field theory, which suggests a way forward.   

In quantum electrodynamics, we already know that exclusive cross sections typically have infrared divergences.   The order-$\alpha$ correction to electron-electron elastic scattering, for example, includes corrections of the general form
\bea
\sigma^{(1)}_{{ee\ra ee}}\left(Q,m_e,{m_\gamma\ne 0},\alpha_{\rm EM}\right) 
&=& \alpha_{\rm EM}\; \beta(Q/m_e)
\  {\ln {m_\gamma\over Q}}
\label{eq:qed-1}
\eea
where $m_\gamma$ is a regulating mass for the photon, without which the correction diverges, $Q$ is the momentum transfer, and $\beta(Q/m_e)$ is a specific function, which itself grows as a log of $Q/m_e$ for $Q\gg m_e$.   So long as $Q/m_e$ is not too large, however, this problem is solved  by introducing a (Bloch-Nordsieck) ``energy resolution", $\epsilon Q$  \cite{blochn}, where we group with true elastic scattering a sum over photons up to the resolution,  ${E_\gamma \le \epsilon Q}$.   Combining these two possibilities, it is found that (\ref{eq:qed-1}) is replaced by
\bea
\overline{\sigma}^{(1)}_{{ee\ra ee}+X(\epsilon)}\left(Q,m_e,{\epsilon Q},\alpha_{\rm  EM}\right) 
&=& \alpha_{\rm EM}\; \beta(Q/m_e)\, {\ln {\epsilon}}
\label{eq:qed-2}
\eea
where at order $\alpha_{\rm EM}$, $X$ denotes one $\gamma$ or nothing.   In contrast to (\ref{eq:qed-1}), this order-$\alpha_{\rm EM}$ correction to the Born cross section is small if $\alpha_{\rm EM}\ln (1/\epsilon)$ is small, and this is the case so long as $\epsilon \gg e^{-1/{\alpha_{\rm EM}}}$, which is a weak condition indeed.   On further reflection we see that any practical experiment will be of this form because of the impossibility of observing arbitrarily soft photons, of which we must have a large number, given the required radiation by accelerated charges in the classical limit.   The finiteness of such soft photon-inclusive cross sections is thus fundamental and motivated by very general considerations.

It is natural to ask whether something like this could be generalized to field theories with massless charged particles, like massless QED, or QCD, whose gluons are both charged and massless.   That is, can we identify observables that have no factors like  $\ln\left ( \frac{m}{Q}\right)$, only at worst $\left ( \frac{m}{Q}\right)\ln\left ( \frac{m}{Q}\right)$?   The answer is yes, but an energy resolution not enough.   An  extra {\it angular} resolution, however, works \cite{ste78,ste77,Sterman:1979uw}.   This result is closely related to the Kinoshita, Lee-Nauenberg Theorem, the infrared finiteness of total transition rates \cite{kln}.

We can bring these abstract considerations into contact with experiment by trading the zero-mass limit for the high-energy limit.    Jet cross sections can then be defined in terms of an angular resolution or its generalizations.  As we shall see below, the essential condition is that final states be ``weighted" by a smooth function of particle momenta, chosen so that the weight for any state with two parallel-moving particles is the same as the weight for the state in which this pair of particles is replaced by a single particle of the same total four-momentum.      Such cross sections are free of infrared divergences of any kind, and depend only on an overall energy scale, $Q$, and a set of dimensionless parameters that define the weight function.   

Cross sections of this kind are perfect for an asymptotically free theory like  QCD, in which $\alpha_s(Q)$ decreases with $Q$.    These include suitably-defined jet cross sections, for which we may shift the renormalization scale to the momentum scale, and write
\bea\sigma\left(Q/\mu,\delta,\alpha_s(\mu)\right)
=
\sigma\left(1,\delta,\alpha_s(Q)\right)\, ,
\eea
where $\delta$ stands for dependence on the parameters (like the angular or energy resolution) on which the weight function depends.
Such cross sections are present in any theory, including the full Standard Model at very high energies $\gg M_W,\, M_Z,\, M_H$ \cite{Butterworth:2008iy}.   

\begin{figure}[htb]
\centerline{%
\includegraphics[width=6cm]{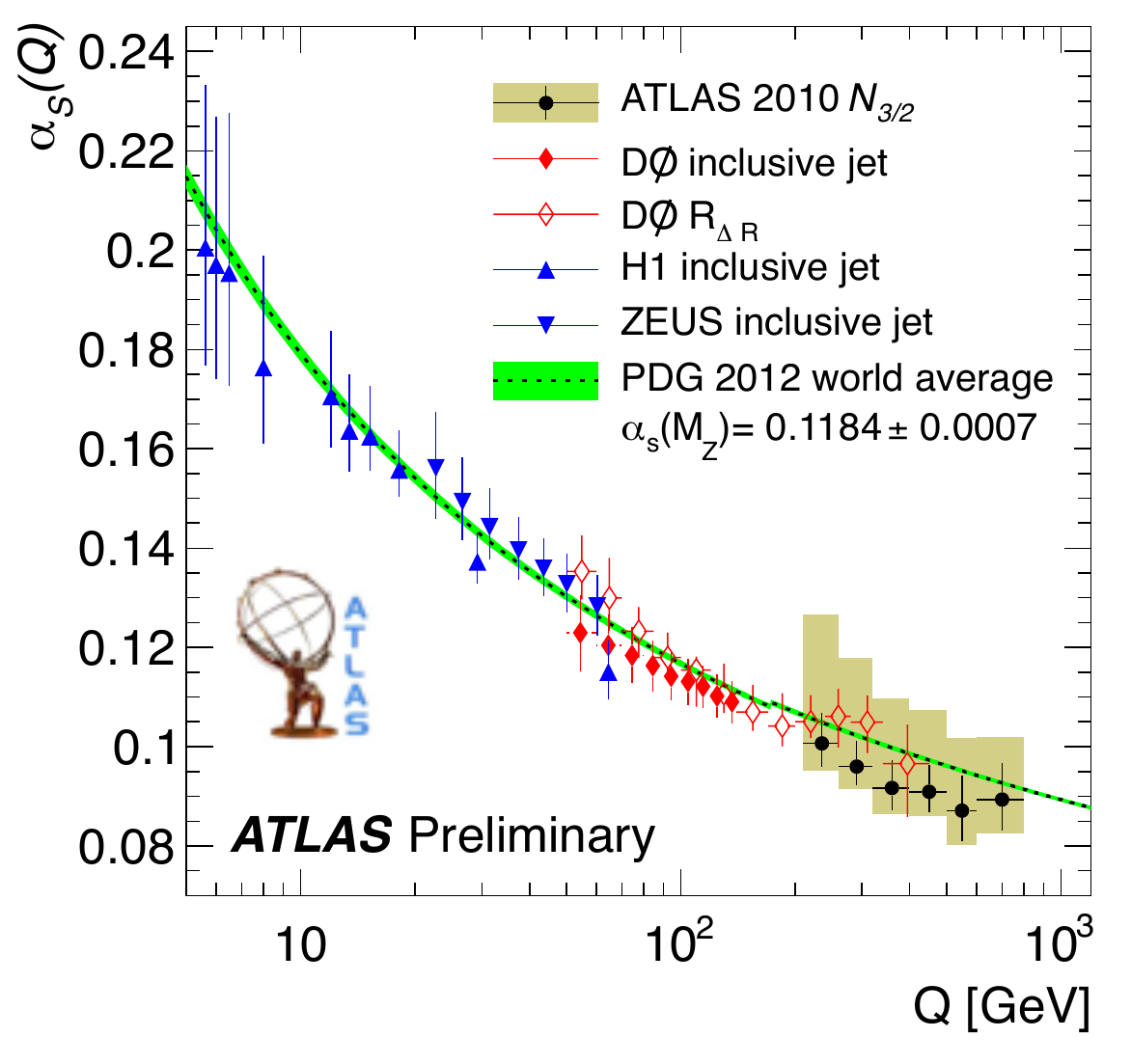}}
\caption{Experimental determination of the running of the coupling.}
\label{fig:atlas-alphas}
\end{figure}

We will review arguments that show how infrared divergences cancel later, but first we discuss how such finite cross sections allow the determination of $\alpha_s$ from an infrared safe cross section.   We suppose we have succeeded in computing a cross section $\hat \sigma$ to some fixed power $n_{\rm max}$ in perturbation theory, and that we have a measurement of this same quantity.   Combining the measured value and the computed expression, we only need solve the equation
\bea
{\hat \sigma}(\alpha_s,Q) = \sum_{n=0}^{n_{\rm max}}
 C_n(Q/\mu)\alpha_s^n(\mu) +\Delta\ &\rightarrow&
\alpha_s(\mu)=f(\sigma(\mu),C_n(Q/\mu),\Delta)\, .
\nn\\
\label{eq:solve-alphas}
\eea
This can be done for any number of observables, with the renormalization scale $\mu$ chosen to minimize the explicit logarithms that occur in the coefficients $C_n$.   Figure \ref{fig:atlas-alphas} shows the result of this procedure applied to many different observables including recent LHC data, and clearly showing the running of the QCD coupling.

Freedom from arbitrarily long-time dependence is called ``infrared safety".   In QCD, confinement involves only low momentum transfers, corresponding to long time scales, and takes place too late to affect the gross properties of energy flow.  The scale for the longest times sampled is set by parameters imposed on the final state, like jet cone sizes or final state jet masses.  A natural conclusion is that any observable based on the flow of energy is infrared safe, since it is unaffected by rearrangements of energy between collinear particles or by soft radiation.  And conversely, probing the details of energy flow (by examining the substructure of jets) through smooth weight functions provides a controllable approach to infrared dynamics.

\begin{figure}[htb]
\centerline{%
\includegraphics[width=10cm]{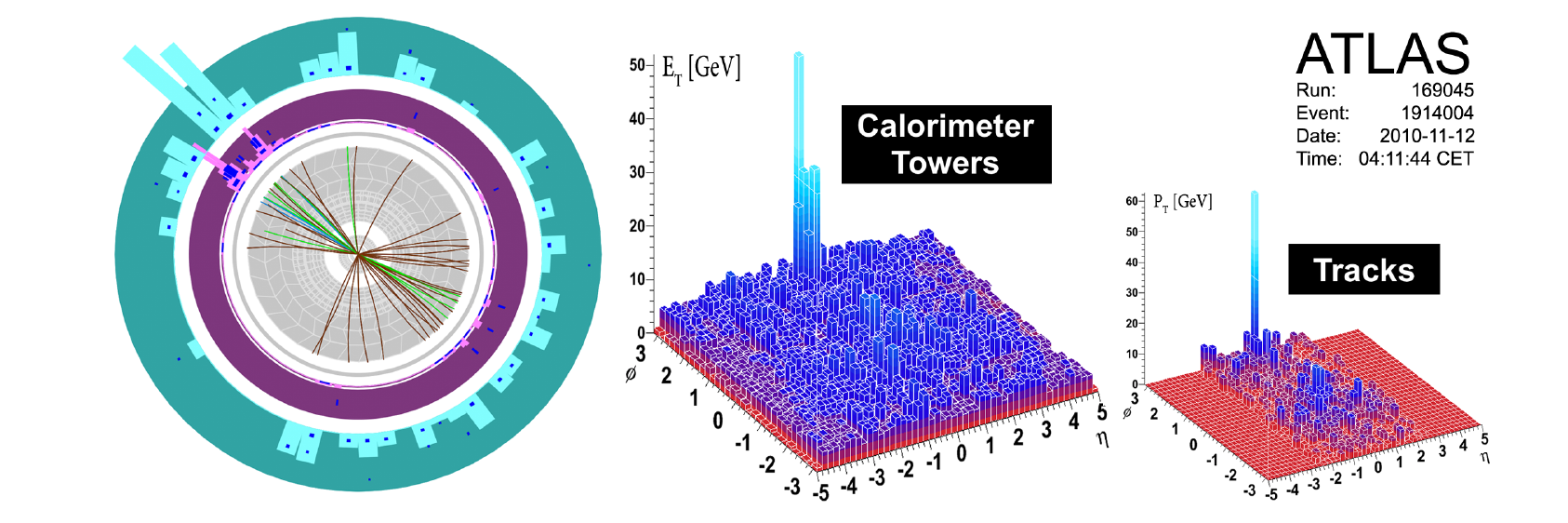}}
\caption{An asymmetric jet event from a nuclear collision at the LHC.}
\label{fig:lhc-AA-jet}
\end{figure}

The collimated nature of jets makes them useful in another realm, where their evolution can be affected by, and carry information from, new phases of strongly-interacting matter in nuclear collisions at RHIC and the LHC.   At both of these colliders the profound effects of the medium on energy flow has been clearly observed \cite{Aad:2010bu}, for example in asymmetric jet pairs like the one shown in Fig.\ \ref{fig:lhc-AA-jet}.

At future electron-nuclear colliders \cite{Accardi:2012qut}, high energy jets and their fragmentation will probe cold nuclear matter, which can shed light both on nuclei and, by comparison with the process in vacuum, on the transformation of partons into hadrons, one of the great challenges of quantum field theory.

\noindent
\section{Subsummary: Lecture I}

The  idea of particle jets from quarks and gluons began haltingly, but has grown into the language of high energy phenomenology.
 Jets are messengers of quantum fluctuations at short distances, capable of bridging  the shortest distances probed at a collider and the size of its detectors.
 Their presence reflects universal behavior in perturbative quantum processes at the confluence of quantum and classical behavior.  
 
 Their dynamical evolution (including absorption) in media provide high-energy probes of those media. They encode the transition from short to long distances, from weak to strong coupling.   In this respect, they are an exemplary 
expression of the gauge theory of color.  Because jets are a general features of quantum field theory, at the very highest energies jets from sources other than QCD may become important. They are integral to the search for new physics at hadron colliders in the production of signals and as background.    

  The second lecture is an introduction to all-orders analysis in perturbative QCD.

\section{Using Asymptotic Freedom with Infrared Safety}

We now return to the role of infrared safety in perturbative QCD, assuming the basic feature of asymptotic freedom, in which the coupling decays as the  renormalization  scale increases.   In any calculation in perturbative QCD at fixed order, we would like to choose the renormalization scale $\mu$ ``as large as possible",  to make $\alpha_s(\mu)$ as small as possible, and thereby make the perturbative expansion as reliable as possible. But how small {\it is} possible?

Consider a ``typical" cross section  (scaled to be dimensionless), and  define $Q^2=s_{12}\equiv (p_1+p_2)^2$ and  $x_{ij}=s_{ij}/Q^2$, where $s_{ij}$ represents the other bilinear invariants that can be made from observed momenta.   Expanded perturbatively, such a cross section will look like
\bea
\sigma\left(\frac{Q^2}{\mu^2},x_{ij},\frac{m_i^2}{\mu^2},\alpha_s(\mu)\right)
=
\sum_{n=1}^\infty C_n\left(\frac{Q^2}{\mu^2},x_{ij},{\frac{m_i^2}{\mu^2}}\right)\, \alpha_s^n(\mu)\, ,
\label{eq:pert-cs}
\eea
with the $m_i^2$ particle masses -- external, quark, and gluon (=0!)   Generically, the order-by-order coefficients $C_n$ depend logarithmically on their arguments, so a choice of large $\mu$ results in large logs of $m_i^2/\mu^2$.   But, as we observed in the first lecture, when we can find quantities that depend on $m_i$ only through powers,  $(m_i/\mu)^p,p>0$, the large-$\mu$ limit exists, and we can write,
\bea
\hspace{-5mm}
\sigma\left(\frac{Q^2}{\mu^2},x_{ij},{\frac{m_i^2}{\mu^2}},\alpha_s(\mu)\right)
&=&
\sum_{n=1}^\infty C_n\left(\frac{Q}{\mu},x_{ij}\right)\, \alpha_s^n(\mu)
+
{\cal O}\left( \left[{\frac{m_i^2}{\mu^2}}\right]^p\right)\, .
\label{eq:use-ire}
\eea
These are the infrared safe quantitates identified above.  Most of the phenomenological side of perturbative QCD depends on the isolation and computation of IR safe quantities.

 To  find IR safe quantities consistently and systematically, we need to understand where the low-mass logs come from.  To analyze diagrams, we generally think of the $m\rightarrow 0$ limit in $m/Q$.   These ratios are the arguments of infrared (IR) logarithms.

\begin{figure}[htb]
\centerline{%
\includegraphics[width=7cm]{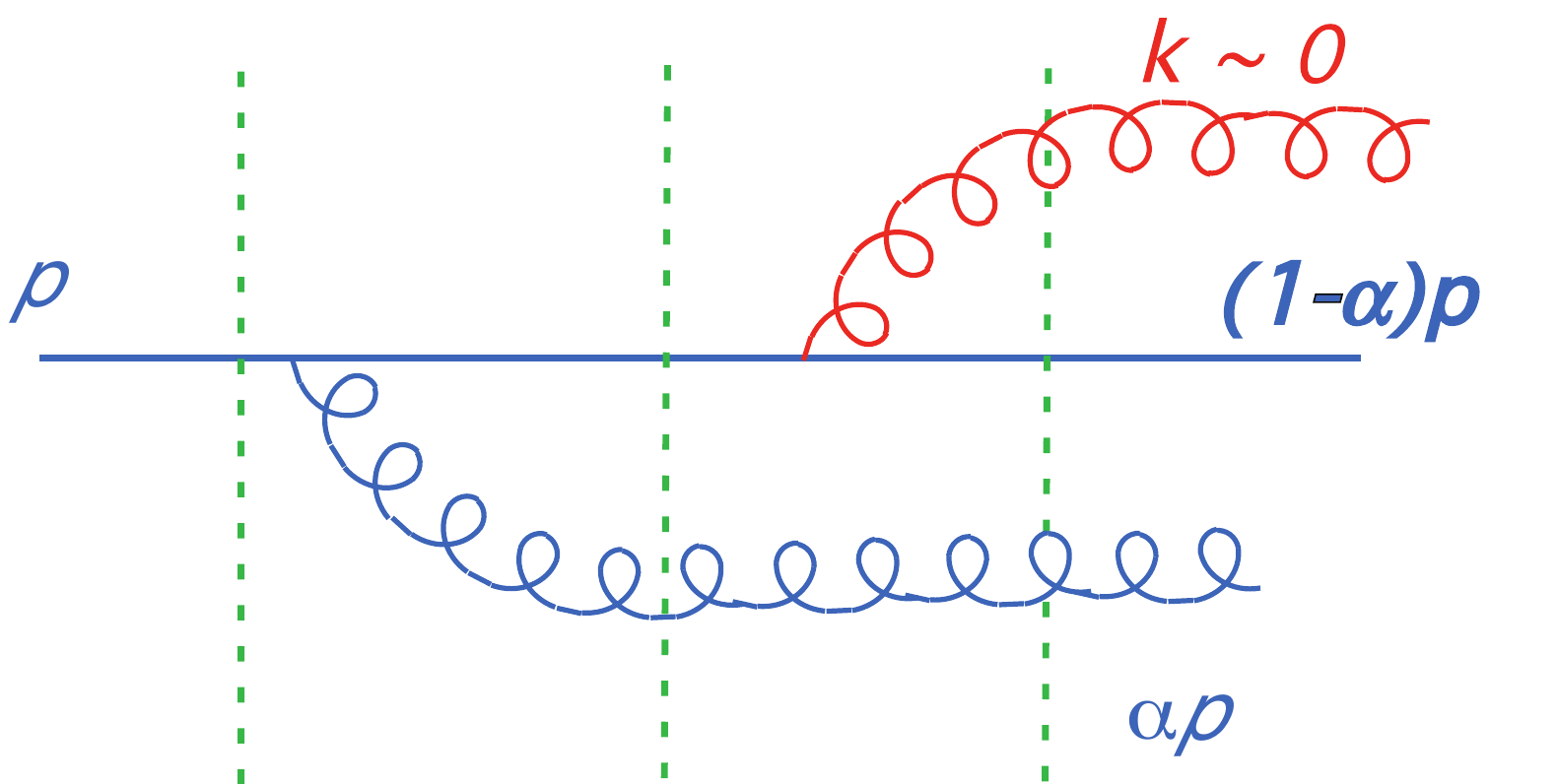}}
\caption{Collinear and soft emission of gluons connecting degenerate states.}
\label{fig:states}
\end{figure}

 As we have seen, the generic sources of IR (collinear and soft) logarithms are illustrated by Fig.\ \ref{fig:states}.
After a short while, noncollinear particles are too separated to interact.   For soft emission and collinear splitting, however, it's ``never too late".   But these
processes don't change the flow of energy.   The connection to energy flow will be confirmed below from a general analysis of Feynman diagrams, and will allow us to prove the IR safety of jet cross sections.

 For IR safety, we want to sum over sets of degenerate final states in perturbation theory, without asking how many particles of each kind we have.  To calculate an IR safe cross section at fixed order, however, the most direct approach is first to compute the cross section for each channel, and then to combine them, although it is worth noting that other approaches are possible \cite{Belitsky:2013xxa,Ore:1979ry}.   Each fixed number of particles has divergences, which should cancel in the sum.    To implement such an approach, we must introduce a new regularization, this time for IR behavior.
The resulting IR-regulated theory is like QCD at short distances, but is better behaved at long distances.   Correspondingly, IR-regulated QCD is {\it not the same as QCD} except for IR safe quantities.  Similar considerations apply to factorized cross sections and amplitudes.   

\begin{figure}[htb]
\centerline{%
\includegraphics[width=8cm]{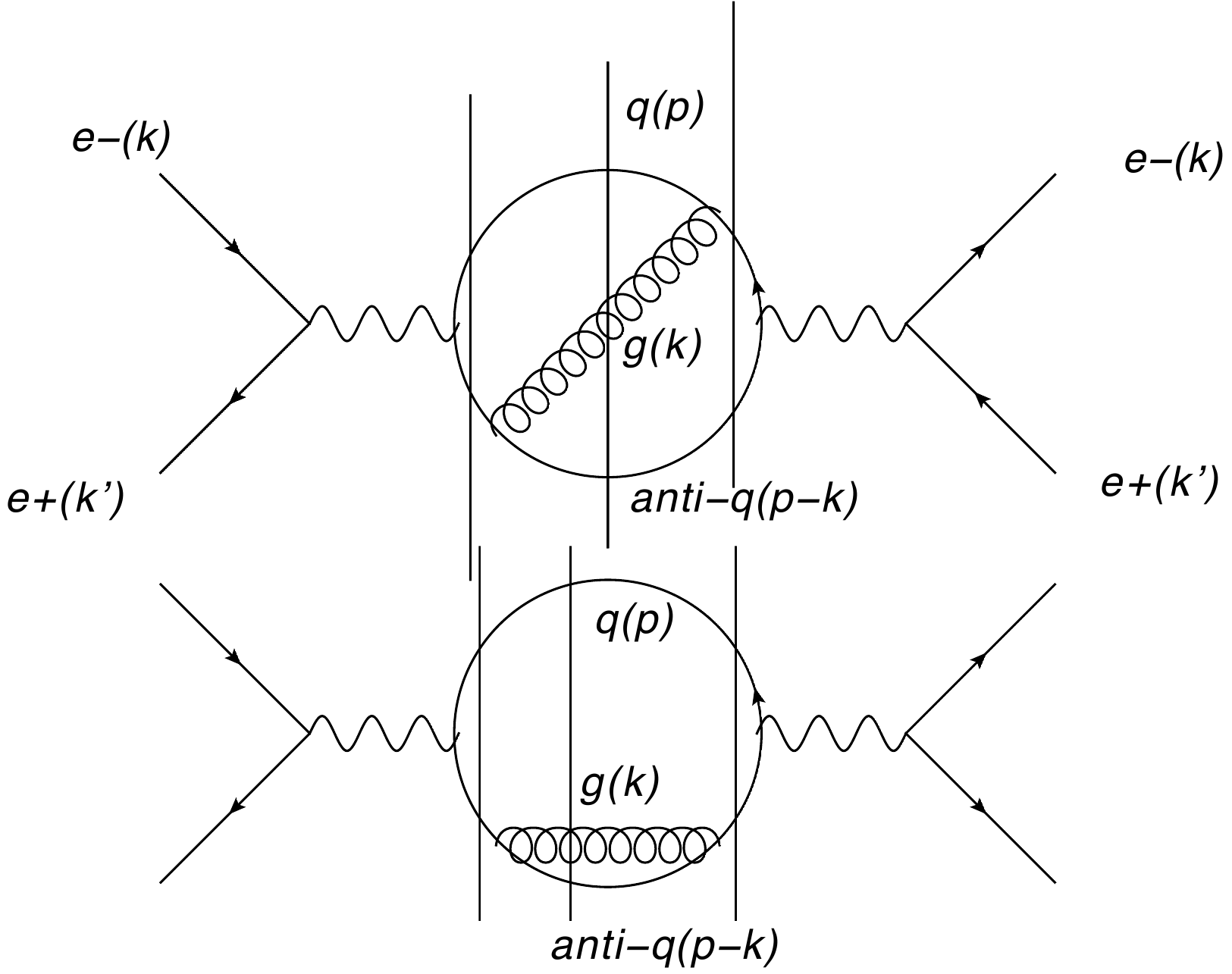}}
\caption{Total ${\rm e^+e^-}$ cross section at order $\as$ in cut diagram notation.}
\label{fig:cut-epem}
\end{figure}

Let's see how this  works for the total $\rm e^+e^-$ annihilation cross section to hadrons (quarks and gluons) at order  $\alpha_s$.    The lowest order, $\alpha^2_{\rm EW}\, \alpha_s^0$, gives the basic a $2\rightarrow 2$ electroweak process, $\sigma_2^{(0)}\equiv \sigma_{\rm LO}$, while the cross section with one-gluon emission, $\sigma_3$, begins at   order $\alpha^2_{\rm EW}\, \alpha_s$, at which order $\sigma_2$ also gets a (virtual) correction.  The order $\as$ cross section is shown in Fig.\ \ref{fig:cut-epem} in a ``cut diagram" organization, in which the amplitude and (time-reversed) complex conjugate amplitude are joined at a vertical line representing the final state.  The uncut diagram is  a forward-scattering process, whose possible cuts enumerate all possible final states.

 The contribution of each final state (cut) to the total cross section is divergent for vanishing quark and gluon masses in normal QCD.   So, we must regularize, then combine final states.   We make two 
representative choices for the regulatization, each corresponding to a modification of QCD.

First we use a gluon mass regularization:  $1/k^2 \rightarrow 1/(k^2-m_g)^2$.   In this case, we find for the cuts of the diagram with two and three-particle final states (up to corrections that vanish as powers of $m_g/Q$),
\bea
\sigma_3^{(m_G)} &=& \sigma_{\rm LO} \frac{4}{3}\frac{\alpha_s}{\pi}\left( 2\ln^2\frac{Q}{m_g}
- 3\ln\frac{Q}{m_g}  -\frac{\pi^2}{6} + {\frac{5}{2}} \right)\, ,
\nn\\
\sigma_2^{(m_G)} &=& \sigma_{\rm LO} \left[ 1 -  \frac{4}{3}\frac{\alpha_s}{\pi}\left( 2\ln^2\frac{Q}{m_g}
- 3\ln\frac{Q}{m_g}  -\frac{\pi^2}{6} + {\frac{7}{4}} \right)\,  \right]\, ,
\label{eq:gluon-mass}
\eea
which gives for the combination, after setting $m_g$ back to zero,
\bea
\sigma_{\rm tot} = \sigma_2^{(m_G)} + \sigma_3^{(m_G)} =
\sigma_{\rm LO} \left[ 1 + {\frac{\alpha_s}{\pi}} \right]\, .
\label{eq:order-alphas}
\eea
This indeed works -- the sum is finite and pretty simple, too.  As expected, there is a cancellation between the virtual ($\sigma_2$)
and real ($\sigma_3$) gluon cross sections.

The second choice is dimensional regularization, which in this context we can think of as a simple change in the area of a  sphere of radius R,
\bea
4\pi R^2 
\Rightarrow (4\pi)^{(1-\vep)}
\frac{\Gamma\left(1-\vep\right)}{\Gamma(2(1-\vep))}
R^{2-2\varepsilon}\, ,
\eea
 with $\varepsilon = 2 - D/2$ in $D$ dimensions, and a substitution in the coupling $g_s \to g_s\mu^\vep$ with $\mu$ the renormalization mass.
 
  Doing the integrals this way, we get 
\bea
\hspace{-80mm}
\sigma_3^{(\varepsilon)} &=& \sigma_{\rm LO} \frac{4}{3}\frac{\alpha_s}{\pi} 
\left( \frac{(1-\vep)^2}{(3-2\vep)\Gamma(2-2\vep)}\right)\, \left(\frac{4\pi\mu^2}{Q^2}\right)^{\vep}
\nn\\
&\ & \times
\left( \frac{1}{\vep^2}
- \frac{3}{2\vep}  -\frac{\pi^2}{2} + {\frac{19}{4}} \right)
\nn\\
\sigma_2^{(\varepsilon)} &=& \sigma_{\rm LO}\, 
\Bigg[ 1 - \frac{4}{3}\frac{\alpha_s}{\pi}
\left( \frac{(1-\vep)^2}{(3-2\vep)\Gamma(2-2\vep)}\right)\, \left(\frac{4\pi\mu^2}{Q^2}\right)^{\vep}
\nn\\
&\ & \times \left( \ \frac{1}{\vep^2}
 - \frac{3}{2\vep}  -\frac{\pi^2}{2} + {4} \right)\,  \Bigg ]\, .
\label{eq:dr}
\eea
Comparing to the gluon mass regularization, although the exclusive channels are quite different, their sum is just the same,
\bea
\sigma_{\rm tot} = \sigma_2^{(\vep)} + \sigma_3^{(\vep)} =
\sigma_{\rm LO}\left[ 1 + {\frac{\alpha_s}{\pi}} \right]\, ,
\label{eq:order-alphas-2}
\eea
where we neglect terms that vanish as $\varepsilon \rightarrow 0$, that is, as we go back to four dimensions.
This example illustrates the principle of IR safety: the specific channels $\sigma_2$ and $\sigma_3$ depend on long time behavior, and therefore on the choice of regulator, but their is sum does not.

 A general rule for  IR safe jet cross sections is to form a  sum over
all states with the same flow of energy into the final state \cite{Sterman:1975xv}. More specifically, we introduce a weight function $e(\{p_i\})$, and define
\bea
\frac{d \sigma}{ d e}\ =\ \sum_n \int_{PS(n)} |M(\{p_i\})|^2 \delta \left ( e(\{p_i\})-e\right)\, ,
\label{eq:weight-sigma}
\eea
where the weight function satisfies
\bea
e(\dots {p_i} \dots p_{j-1},{\alpha p_i},p_{j+1}\dots ) \ =\ e(\dots {(1+\alpha) p_i} \dots p_{j-1},p_{j+1}\dots )\, .
\label{eq:weight-properties}
\eea 
 We will neglect long times in the initial state for the moment,  and see how this works in $\rm e^+e^-$ annihilation
jet cross sections.

\section{Infrared Safety to All Orders for Cross Sections: Jets and Event Shapes}

\subsection{Cross sections, cut diagrams and generalized unitarity}

\begin{figure}[htb]
\centerline{%
\includegraphics[width=10cm]{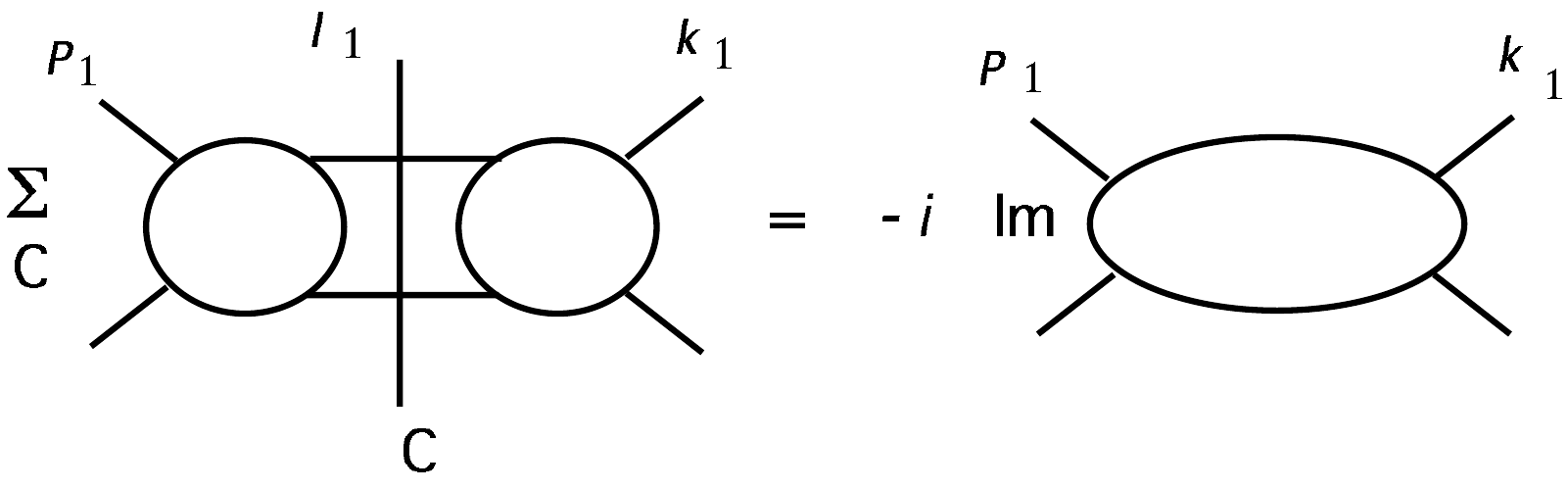}}
\caption{Unitarity.}
\label{fig:cut-unitarity}
\end{figure}

\begin{figure}[htb]
\centerline{%
\includegraphics[width=10cm]{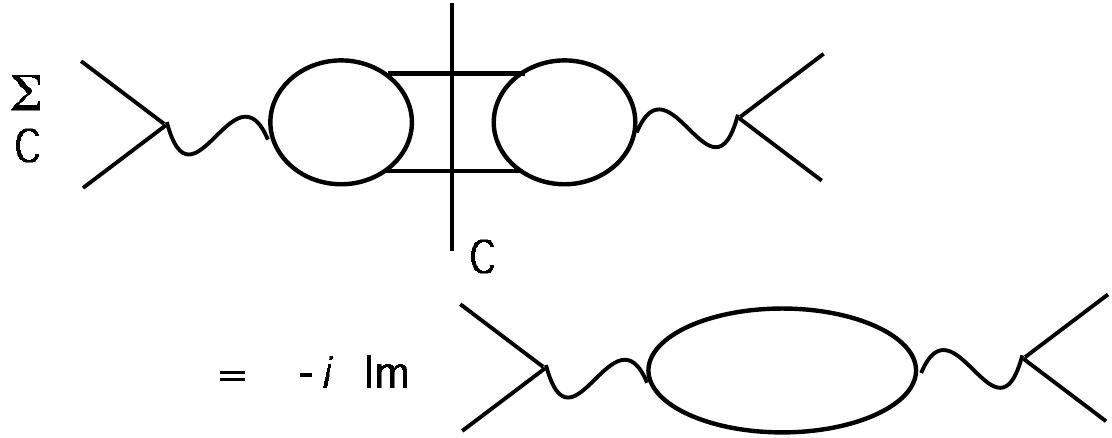}}
\caption{Total ${\rm e^+e^-}$ cross section in cut diagram notation.}
\label{fig:epem-allorder}
\end{figure}

It is possible to prove the infrared safety of many jet-related cross sections to all orders, starting from very general properties of perturbation theory.  One of the most basic of these is unitarity, illustrated for a general $2\to 2$ process in Fig.\ \ref{fig:cut-unitarity},
\bea
\sum_{{\rm all}\ C}G_C(p_i,k_j\, )
=
2\; {\rm Im}\; \left [-iG(p_i,k_j\, )\right ]\, .
\label{eq:optical}
\eea
This relation is true diagram-by-diagram at each order, and when $\{p_i\}=\{k_i\}$ are two-particle states, it relates the cross sections found from each cut to the forward scattering amplitude.  (The relation holds even when they are multiparticle states and are not equal.)  Now to lowest order in electroweak interactions, the cross section $\rm e^+e^- \rightarrow$ hadrons is given by the cuts of a specific set of diagrams shown in Fig.\ \ref{fig:epem-allorder}, in which the hadrons appear in cuts of the self energy diagram of a photon or Z in the Standard Model.
 
The all-order cross section for inclusive annihilation and decay can thus be reinterpreted in terms of the cut self-energies, as
\bea
 \s_{e^+ e^-}^{(\rm tot)} (q^2) = {e^2 \over q^2} {\rm Im}\, \p (q^2)\, ,
\label{eq:optical-pi}
\eea
where the function $\pi$ is defined in terms of the two-point correlation function of the relevant electroweak currents $J_\m$ (with their couplings included) as
\bea
\hspace{-10mm}
\p (q^2) (q_\m q_\n - q^2 g_{\m\n} ) =  i \int d^4x\, e^{i q x} < 0 |\; T\; J_{\m} (x) J_\n(0)\; | 0 >\, .
\label{eq:pi-of-q}
\eea
We conclude that if we can show that $\p(q^2)$ is IR safe, so will be $\s_{e^+e^-}$, because the latter is proportional to the imaginary part of the former.

 At this point, it's easy to check the IR safety of $\p(q^2)$, by appealing to the necessary condition for sensitivity to long-time behavior: the possibility finding physical processes involving the classical propagation of on-shell lines between the initial and final states.   For $\pi(q^2)$, the initial and final states are the same, with all hadronic energy flowing in from, and  then flowing out to, an electroweak current.  For $q^2>0$, there is simply no physical propagation of massless lines that can carry the total energy from one such current to the other, simply because for $q^2>0$ the massless particles emitted from the action of the current must travel in different directions at the speed of light, and so can never meet at a later point.   These simple considerations are, as we have seen, the origin of jet behavior in final states.  The same reasoning thus shows that in the total cross section the details of jet behavior are forgotten.    

 In fact, the combination of unitarity and physical propagation is much more general, and allows us to prove the IR safety of the weighted cross sections defined by Eqs.\ (\ref{eq:weight-sigma}) and (\ref{eq:weight-properties}).   We do this by generalizing the optical theorem of Eq.\ (\ref{eq:optical}) and Fig.\ \ref{fig:cut-unitarity} to fixed values of the internal spatial loop momenta, $\vec l_a$ of each forward-scattering diagram in the figure,
\bea
\sum_{{\rm all}\ C}G_C(p_i,k_j,\vec l_a\, )
=
2\; {\rm Im}\; \left [-iG(p_i,k_j,\vec l_a\, )\right ]\, .
\label{eq:unitarity-generalized}
\eea
That is, unitarity holds not only for the complete diagram, but also for the diagram when only the energy components of loop integrals are integrated.   (By the way, an analogous relation holds for integrals over a single light cone momentum, $l^0\pm l^3$.)    The proof of this relation, is surprisingly simple.   Doing the time integrals for a general amplitude in Eq.\ (\ref{eq:topt})  gives  the standard form of ``old-fashioned" time-ordered perturbation theory.    For each ordering of the vertices, the time integrals give an overall energy-conservation delta function times a product of energy deficit denominators.    The denominator for each intermediate state $j$ is the energy that has flowed into the diagram before $j$ minus the sum of the on-shell energies of the lines in state $j$.   Summed over the (many) possible time orderings of vertices, this result is completely equivalent to the Feynman diagram with the same set of lines and vertices.   For each time ordering of the diagram $G$ at fixed loop momentum, the sum of cuts in Eq.\ (\ref{eq:unitarity-generalized}) gives a specific sum over final states.   Suppressing overall factors, and fixing spatial loop momenta, we can represent this result as
\bea
\sum_{C=1}^A G_C &=& \sum_{C=1}^{A}
\prod_{j=C+1}^{A}{1\over E_j-S_j-i\ep}
\ (2\p)\delta(E_C-S_C)\prod_{i=1}^{C-1}{1\over E_i-S_i+i\ep}
\nn\\
\nn \\
 &=& -i\left [ -\prod_{j=1}^{A}{1\over E_j-S_j +i\ep} + \prod_{j=1}^{A}{1\over E_j-S_j-i\ep}\right ]\, ,
%\nn
\eea
where $A$ is the number of intermediate states (one less than the number of vertices in the diagram).    The second line of this relation follows very easily by repeated use of the identity
\bea
i\left ({1\over x+i\ep}-{1\over x-i\ep} \right ) = 2\p \delta(x)\, .
%\nn
\eea
We conclude that the sum over the cuts of any diagram even at fixed spatial loop momenta gives a result that is the imaginary part of a forward-scattering amplitude.   The same reasoning as for the total cross section, based on physical propagation, then shows the cancellation of long-distance behavior in the sum over cuts at fixed loop momenta.   This can be applied to the neighborhood of any set of loop momenta which gives long-distance behavior in each individual final state, because when virtual lines become parallel to lines in the final state, these same lines appear in final states of other cuts of the same diagram at these fixed loop momenta, and hence in the same jet of particles with the same total energy.    At fixed loop momenta, the weights associated with Eq.\ (\ref{eq:weight-properties}) will be the same for all states related in this manner, and we can freely carry out the sum of Eq.\ (\ref{eq:unitarity-generalized}).  Only the contributions from virtual particles that are not collinear to those in the final state will be left over, and, as we have argued, these will finite.  This is the basis of the infrared safety of jet cross sections.

\subsection{Jets and event shapes}

There are quite a wide variety of weights and jet cross sections.   The simplest conceptually are the cone definitions, although choosing a fixed cone in space would leave out too many events.   In practice, one first identifies the jets, then associates with them IR safe quantities.   

\begin{figure}[htb]
\centerline{%
\includegraphics[width=5cm]{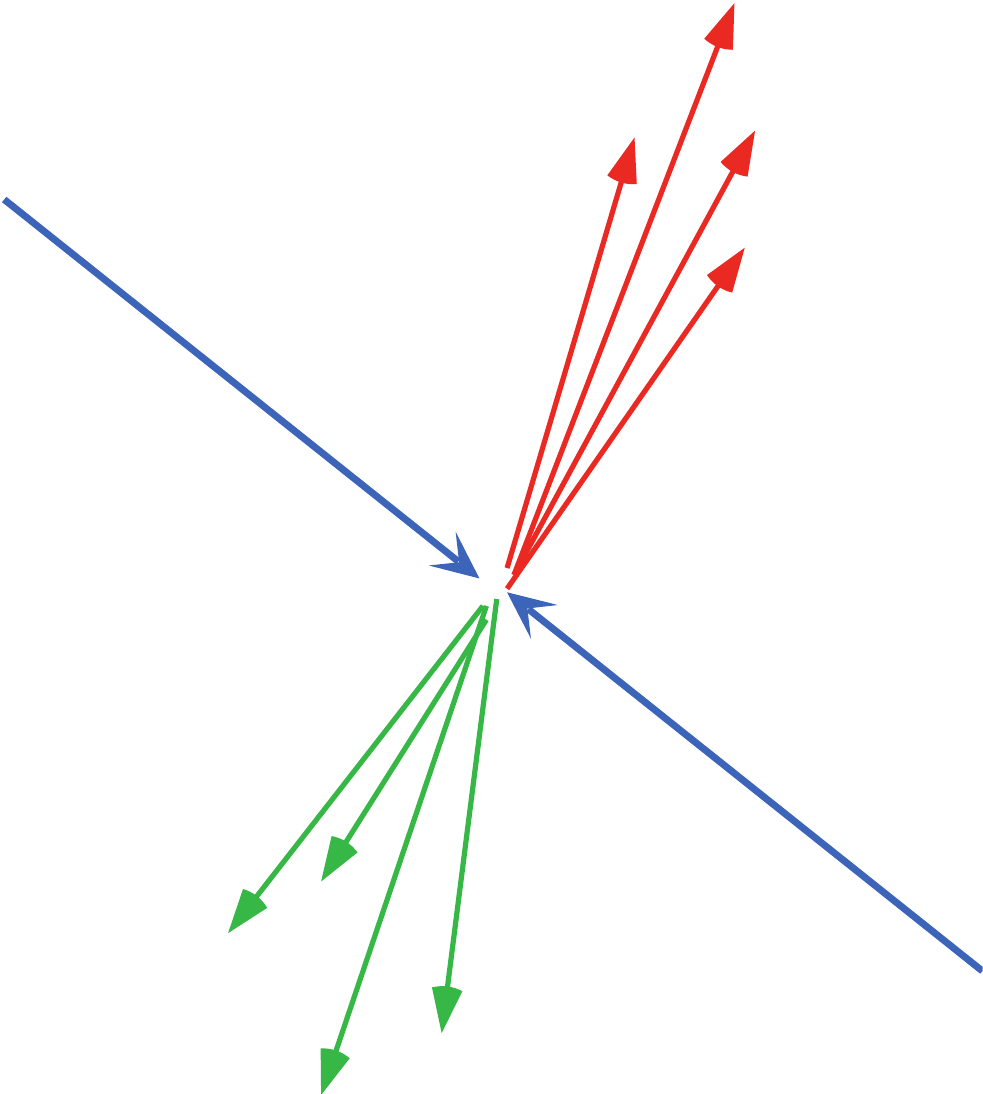}}
\caption{Generic two-jet configuration in the ${\rm e^+e^-}$ annihilation rest frame, naturally characterized by the thrust variable.}
\label{fig:2jet}
\end{figure}

The ``thrust" \cite{Farhi:1977sg} is one of the oldest and most exemplary of the ``event shape variables", which are IR safe variants of jet cross sections \cite{Dasgupta:2002dc}.  It is adapted to the back-to-back jets associated with the quark pair produced in ${\rm e^+e^-}$ annihilation, where the lab frame is the rest frame of the system.   It is designed to identify the axis that best describes the final state as a pair of independently hadronizing jets, moving in opposite directions in the lab frame, as in Fig.\ \ref{fig:2jet},
\bea
T= {1\over Q}{\rm max}_{\hat n}\ \sum_i |{\hat n}\cdot {\vec p}_i|\, ,
\label{eq:thrust-def}
\eea
with $Q$ the center of mass energy.   We easily check that this event shape satisfies the requirements for IR safety, Eq.\ (\ref{eq:weight-properties}).

At energies high enough to neglect all particle masses, $T=1$  characterizes ``back-to-back" jets, each consisting of sets of perfectly collinear particles.  Defining
 $e_T\equiv 1-T$,  we find a relation between $e_T$ and the hemisphere invariant masses,
\bea
e_T\  \sim \ {M_R^2+M_L^2\over Q^2}\, .
\label{eq:e_T}
\eea
The thrust then, is directly sensitive the masses of the jets.    This suggests that we can ``design" event shapes to test other properties of jets.  One such set of generalizations  is found by choosing   $\hat z$ as the thrust axis and dividing the event into right and left $``H_R"$ and $``H_L"$ hemispheres, in terms of which
\bea
e_T = {1\over Q} \left[ \sum_{i\in H_R}\ k_i^-  + \sum_{i\in H_L} k_i^+ \right]
= {1\over Q}\sum_{{\rm all}\ i}\ k_{iT}\; \e^{-|\eta|}\, ,
\eea
with $\eta$ the rapidity of the (massless) particles.
We then define \cite{Berger:2003iw}
\bea
e_a 
= {1\over Q}\sum_{{\rm all}\ i}\ k_{iT}\; \e^{-|\eta|(1-a)}\, .
\label{eq:angularity}
\eea
The parameter $a$ defines a family of event shapes, called angularities \cite{Berger:2004xf} where  $e_0=e_T$, and  $e_1$ is called ``jet broadening" \cite{Catani:1992jc}.

Going beyond two jets in electron-positron annihilation, it is desirable to have algorithms that identify jets even in complicated events like the one shown in Fig.\ {\ref{fig:cms-jet}.    These ``cluster" or ``jet finding" algorithms play in more general kinematic situations the role played by the maximization in the definition of thrust, Eq.\ (\ref{eq:thrust-def}).  These algorithms form jets by successively combining particles (or calorimeter cells) by identifying the minimum of a ``cluster variable" among all pairs, combining that pair into one particle or cell, and then proceeding until all resulting pairs exceed a certain value of the cluster variable.   The cutoff value is analogous to a cone or energy resolution.   The final set of combined particles or cells defines a list of jets.     For hadronic collisions, the favored cluster variables are naturally built around momenta $k_{ti}$, transverse to the beam directions,
\bea
d_{ij} = {\rm min}\, \left( k_{ti}^{2p}\, ,\,  k_{tj}^{2p} \right)\, \frac{\Delta^2_{ij}}{R^2}\, ,
\eea
where $\Delta_{ij}^2 = (y_i-y_j)^2 +(\phi_i-\phi_j)^2$ in terms of rapidities $y_i$ and azimuthal angles, $\phi_i$. $R$ is an adjustable
parameter.
 The ``classic" choices are $p=1$, the ``$k_t$ algorithm,  $p=0$ the ``Cambridge/Aachen" algorithm, and $p=-1$, the  ``anti-$k_t$" algorithm \cite{Cacciari:2008gp}, which in practice comes close to identifying  jets made up of cone-like ensembles of particles.

\section{Factorization, Evolution and Resummation}

We have seen that it is independence of long-distance behavior that makes jet cross sections infrared safe when, as in ${\rm e^+e^-}$ annihilation, all partons emerge from a single point in space-time.    More specifically, we concluded that infrared safe observables should depend on only the flow of energy into the final state, which ensures independence of collinear re-arrangements and soft parton emisssion.    But in collisions involving hadrons, we {prepare} one or two particles in the initial state (as in DIS or proton-proton scattering), particles which themselves are obviously  sensitive to long time behavior of QCD, since this is what binds them in the first place.   
The parton model suggests what to do in such cases: factorize.    That is, try to separate the cross section into short-distance parts that we can confirm are infrared safe, multiplied by, or in convolution with, long-distance functions that we can determine from experiment.    We have already illustrated this form in Eqs.\ (\ref{eq:jet-fact}) and (\ref{eq:1pi}) for jet and single-particle incusive cross sections.   The universal parton distributions take into account, for example, the soft  fragments seen in the original cosmic ray ``jets"  mentioned at the beginning of these lectures.

The general form we seek is
\bea
Q^2\sigma_{\rm phys}(Q,m)
&=&
{\omega_{\rm SD}(Q/\mu,\as(\mu))}\, \otimes\, f_{\rm LD}(\mu,m) 
+ 
{\cal O}\left(\frac{1}{ Q^p}\right) ,
\label{eq:fact-general}
\eea
where $p>0$, $\otimes$ is a convolution as in Eqs.\ (\ref{eq:jet-fact}), (\ref{eq:1pi}), and where $\mu$   is the  factorization scale, which we must introduce to separate what we choose to call long- and short-distances (``LD" and ``SD").   The default choice of $\mu$ is the hard scale, $Q$, assuming, of course, that there is a unique hard scale.  Note that $\mu$ appears only on the right-hand side of this expression, not the left, which represents a physical quantity.    The parameter $m$, which appears on both the right and left, represents physical mass scales that we choose to be on the IR side of $\mu$.   Sometimes (especially for jet cross sections) $m$ may include perturbative scales, as well as fully infrared scales like light parton masses.   

If there is heavy-particle ``new physics", associated with very short-lived virtual states, it can be incorporated  in  $\omega_{\rm SD}$, while $f_{\rm LD}$ incorporates ``universal" corections associated especially with the structure of the colliding hadron(s).  Equations of the form of (\ref{eq:fact-general})  are the basic input to almost all collider applications.     As in the case of the truly IR safe jet cross sections of ${\rm e^+e^-}$, factorization, Eq.\ (\ref{eq:fact-general})  requires a smooth weight for final states, Eq.\ (\ref{eq:weight-properties}).   The actual calculations involve an infrared regularization of the kind discussed in connection with Eqs.\ (\ref{eq:gluon-mass}) and (\ref{eq:dr}).  We calculate the right-hand side of (\ref{eq:fact-general})  and the universal functions $f_{\rm LD}$ in the regulated theory, which enables us to derive the short-distance coefficients $\omega_{\rm SD}$.    The calculation, assumes that we can construct $f_{\rm LD}$ order-by-order in IR-regulated perturbation thoery.  For details, see, for example, Refs.\ \cite{Brock:1993sz,Sterman:1995fz,Collins:2011zzd}.     With $\omega_{\rm SD}$ in hand, we discard the regulated theory, and rely on a library  of  experimentaly-determined parton distributions $f_{\rm LD}$ at appropriate values of $\mu$, to provide predictions for  the cross sections on the left-hand side of (\ref{eq:fact-general}). This is a self-consistent ``bootstrap" process, in which the experimental parton distributions themselves are determined by comparing measured cross sections on the left of (\ref{eq:fact-general}) with computed short-distance coefficients for sets of benchmark processes.    Of course, no finite number of measurements can determine $f_{\rm LD}$ for all $\mu$, and that is where the technique of evolution, which can be thought of as a consequence of factorization, comes in.

 Whenever we succeed in writing a factorized expression for a cross section we can derive an equation for its dependence on the factorization scale $\mu$, because the cross section, which is a physical quantity, cannot depend on our choice of the renormalzation scale,
\bea
0=\mu{d\over d\mu} \ln \sigma_{\rm phys}(Q,m)\, .
\label{eq:mu-indep}
\eea
Then, by invoking separation of variables, we have, schematically,
\bea
\mu{d \ln f\over d\mu}= - P(\as(\mu)) = - \mu{d \ln \omega \over d\mu}\, ,
\label{eq:separate}
\eea
where $P(\as)$ can depend only on the variables held in common by the short- and long-distance factors.  Aside from $\as$, these are the momentum fractions $x$ and $z$ of Eqs.\ (\ref{eq:jet-fact}) and (\ref{eq:1pi}), which we suppress here. Following this reasoning, wherever there is factorization, there is an ``evolution equation", whose solution is a form of ``resummation", which is simply the solution to the corresponding equation \cite{Contopanagos:1996nh}.   In the simplest case of a product, we have
\bea
 \sigma_{\rm phys}(Q,m) = \sigma_{\rm phys}(q,m)\ \exp\left\{  \int_q^Q {d\mu'\over \mu'} 
P\left( \alpha_s(\mu')\right) \right\}\, .
\label{eq:solution}
\eea
 The beauty of this approach is that the ``separation constant" $P(\alpha_s)$ can be computed from the short distance function, and so is IR safe, and at the same time determines the $\mu$-dependence of the long-distance functions, allowing us to use data from relatively low momentum transfers to make predictions at high momenta.   This is how predictions can be made for the LHC jet cross sections of Figs.\ \ref{fig:jet-sigma} and \ref{fig:jet-theory-to-exp}, making use principally of the celebrated DGLAP evolution for parton distributions \cite{dglap}.     In addition, resummation sometimes makes possible predictions in limits that are not infrared safe order-by-order in perturbation theory \cite{Parisi:1979se}

\section{The Spirit of Factorization Proofs}

Space does not allow more than a sketch of proofs of factorization, but we have assembled sufficient concepts in these lectures to do at least this.   In particular, it may be interesting to see the quantum field theoretic analogs of the classical arguments for the ``autonomy" of jet functions given above, which is an essential ingredient in the factorized jet and single-particle cross sections of Eqs.~(\ref{eq:jet-fact}) and (\ref{eq:1pi}).    The technical problem is the factorization and cancellation of soft gluons that, on a diagram-by-diagram basis, attach outgoing jets with each other and with the forward-moving fragments of the colliding hadrons.   The solution to this problem \cite{Libby:1978qf,Collins:1981ta,Feige:2014wja} can be organized into two steps.

The first step is represented by  Fig.\ \ref{fig:frag3}, which shows the sources of long-distance behavior in an arbitrary momentum-space configuration  that contributes to the single-particle inclusive cross section for a hadron, $h(p)$ at some fixed momentum.   The figure shows jet-like subdiagrams, one a final-state jet consisting of lines in the direction of $h(p)$, along with other final-state jets, one coming from the hard-scattering, $H$, and two others consisting of the fragments of the colliding hadrons, labeled $J_1$ and $J_2$.  A dashed oval, labelled $H_{h(p)}$ surrounds all lines in the final state that are not in the forward ($J_i$) directions.  The blob labelled $S$ consists of soft gluons, interacting with each other and with each of the jet diagrams, through propagators of the vector potential $A^\mu$.   We should note that the arguments we have given above are adequate to show that the only connections between the final state jets are through soft gluons.

The right-hand side of Fig.\ \ref{fig:frag3}  illustrates the field-theoretic analog of the classical argument on the Lorentz contraction of the gauge field.   The dynamics of the jet of lines collinear to $p$ factorizes completely from the soft gluons.   While we cannot give details here \cite{cssrv,Sachrajda:1978ja}, the essential point is that in perturbation theory, we can show that the leading-power connections of all gluons on the right-hand side involve not the full field, but only its divergence, very much as in the classical electromagnetic field, Eq.\ (\ref{eq:gauge-field}).   Summing over gauge-invariant sets of diagrams then leads to the right-hand side, which represents the result that the soft gluons couple only to the net color charge of the outgoing parton that fragments into the observed hadron, $h(p)$, but has no influence on the fragmentation itself.   The result is very much of the form of Eq.\ (\ref{eq:1pi}), with a fragmentation function, represented by the diagram on the far right, multiplying the remaining cross section.   
\begin{figure}[htb]
\centerline{%
\includegraphics[width=13cm]{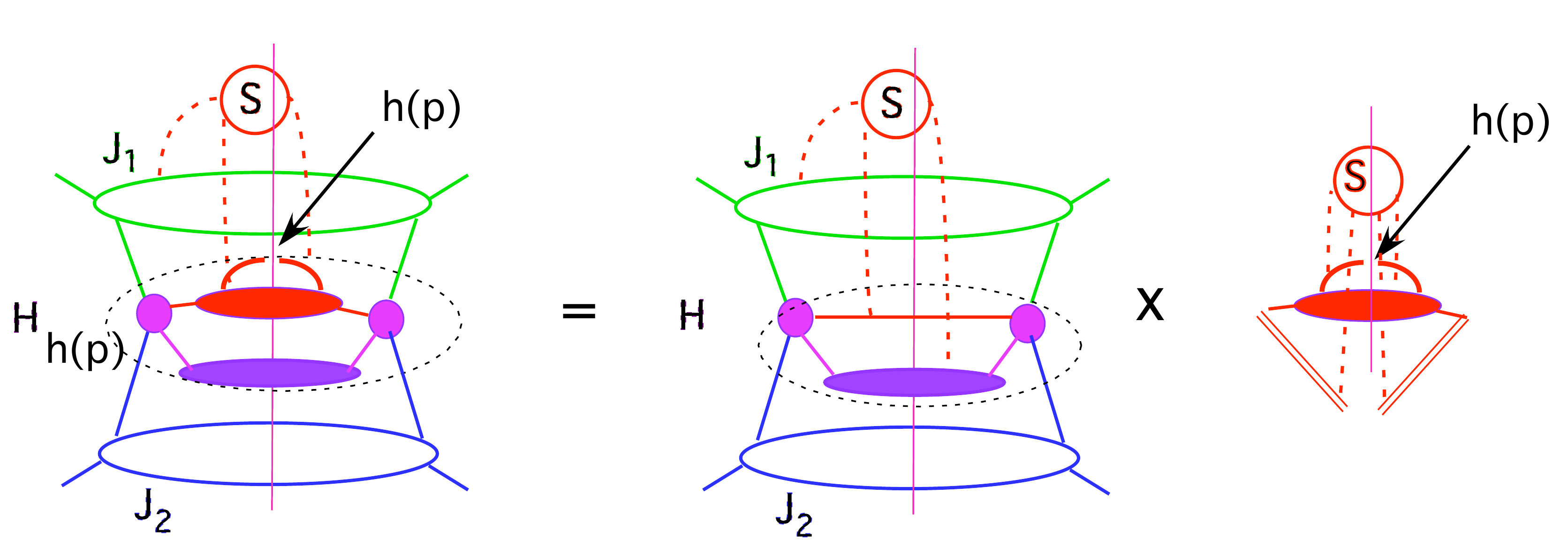}}
\caption{Factorization for single-particle inclusive cross section in cut diagram notation.}
\label{fig:frag3}
\end{figure}

The second step is to show the cancellation of the remaining soft connections to the simplified hard scattering, $H$ on the right-hand side of Fig.\ \ref{fig:frag3}.   This separates the other final-state jets from the forward-moving jets, $J_1$ and $J_2$.   This part of the all-orders argument is another application of the optical theorem, very much in the spirit of our proof of the IR safety of jet cross sections.   The procedure is represented in Fig.\ \ref{fig:LS}.    The first line is the optical theorem applied to the hadron-hadron collision, in the generalized form of Eq.\ (\ref{eq:unitarity-generalized}).   The physical cross section is on the left-hand side, but unlike ${\rm e^+e^-}$ annihilation, Fig.\ \ref{fig:epem-allorder}, in this case we do not have every cut of the diagram, because soft gluons exchanged {\it before} the hard scattering in the amplitude or its complex conjugate can produce states that do not contribute to the jet or single-particle inclusive cross sections, which require that a hard scattering has taken place.  For each term on the right-hand side of the first line, however, there are closed loops of energetic lines that connect hard scatterings.    The same physical picture requirement that we applied to Fig.\ \ref{fig:epem-allorder} shows that these loops are essentially point-like, as shown in the second line.   But then,  all terms on right-hand side are power-suppressed by the scale of the large momentum transfer, because soft radiation is very unlikely to be emitted by lines that are far off-shell.
\begin{figure}[htb]
\centerline{%
\includegraphics[width=13cm]{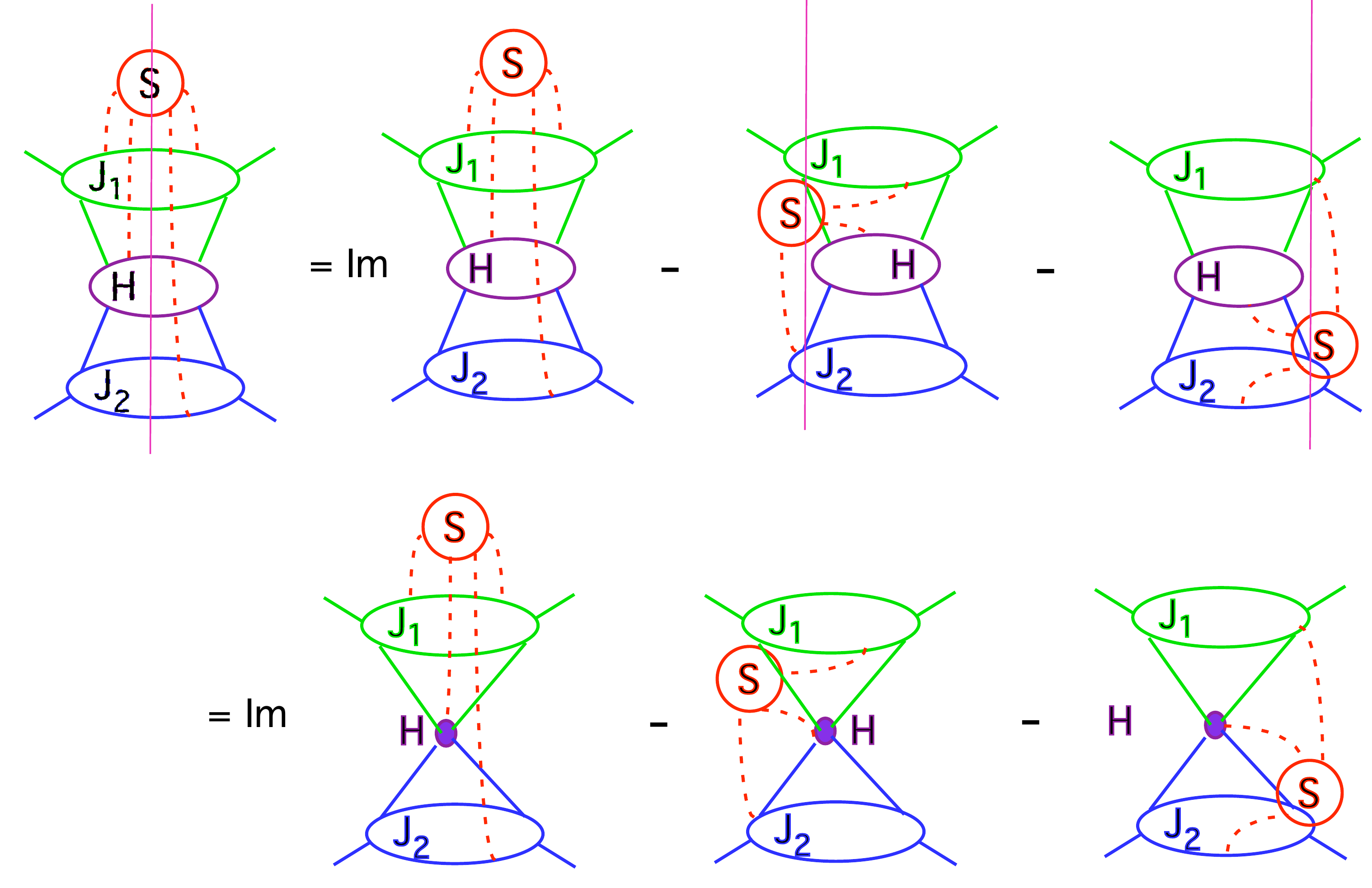}}
\caption{Cancellation of soft gluons that connect the initial state to the final state.}
\label{fig:LS}
\end{figure}
With this result, the factorization of the cross section, Eq.\ (\ref{eq:1pi}) is almost complete, except for soft gluons that do {\it not} connect to the final-state jets, but only connect the forward jets, $J_1$ and $J_2$.  Such connections are also present in processes that depend on electroweak rather than hard scattering, such as the Drell-Yan process.   Proofs of factorization for these processes, and therefore for jet and related cross sections, have been presented in detail in Refs.\ \cite{Bodwin:1984hc,Collins:2011zzd}.

\section{Conclusion}

We have, of course, only scratched the surface of the tremendous amount of work that has gone into perturbative QCD at short distances.  In particular, the extraordinary progress in explicit calculations at fixed order \cite{Hoche:2013zja} has been passed over, although I hope the motivation underlying these calculations has been touched upon.

Another frontier beyond the discussion given above is implied by the structure of factorization proofs outlined in the previous section.   The cancellation of the soft radiation that would connect the dynamical evolution of final- and initial-state jets assumes that we sum over soft radiation of all energies, without imposing an ``energy resolution" $\epsilon$ that is``too small".    The dependence of jet cross sections on the analog of the energy resolution in Eq.\ (\ref{eq:qed-2}) is quite complicated however
\cite{Forshaw:2012bi,Catani:2011st}, and is subtle to control even for relatively simple processes like jets in ${\rm e^+e^-}$ annihilation \cite{Dasgupta:2001sh}.    Some of these subtleties are illustrated by Fig.\ \ref{fig:non-global}, in which we trigger on two jets, with energies of order $Q$, and measure the energy that is radiated into some region $\Omega$ outside the jets.    We denote the distribution of energy flow into $\Omega$ by $\Sigma_\Omega(E)$.
\begin{figure}[htb]
\centerline{%
\includegraphics[width=7cm]{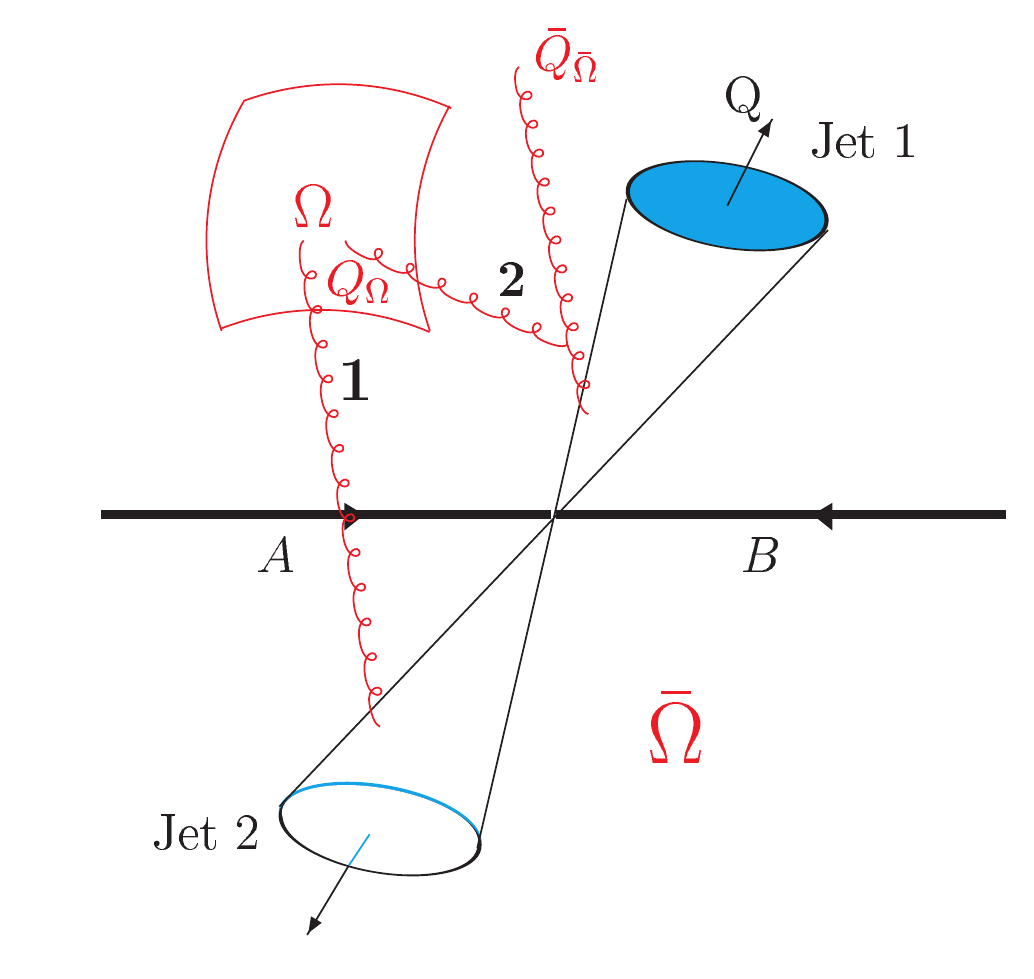}}
\caption{Radiation into a specified region.}
\label{fig:non-global}
\end{figure}

The nature of the cross section depends sensitively on how the intermediate region $\bar \Omega$  is treated.   In brief, if we limit radiation into $\bar\Omega$ ($Q_{\bar\Omega}$ in the figure) enough so that the number of jets is fixed at two, then we can retain a simple form of factorization with only two final-state jets.   In this case, we can derive relatively simple evolution equations for $\Sigma_\Omega(E)$, and resum logarithms of $E/Q$ \cite{Berger:2003iw,Dokshitzer:2003uw}.  The other possibility, which is much less restrictive, is to allow the cross section to be inclusive in region $\bar\Omega$.   In this case, the number of jets is effectively not fixed, as jets of intermediate energy populate region $\bar\Omega$ and radiate into $\Omega$.   Then we cannot write a factorized form for the cross section that is as ``simple" as Eq.\ (\ref{eq:jet-fact}), for example.   It turns out, however, that in an appropriate approximation (large numbers of colors, $N_c$), it is possible to derive a nonlinear evolution equation for $\Sigma_\Omega(E)$ \cite{Banfi:2002hw}, which is nearly identical to the nonlinear equations that describe high energy saturation physics and nuclear scattering \cite{Balitsky:1995ub,Kovchegov:1999yj,Stasto:2004rm} in similar approximations.   The origin of the nonlinearity is the production of color dipoles, which extend out of the two original jets into intermediate region $\bar\Omega$.    Considerations of this sort are becoming more and more relevant for studies of the long-distance behavior of  QCD, for its own sake and for its role as a precision tool in tests of the Standard Model \cite{Berger:2010xi} and the search for New Physics.   

In summary, we have a good understanding of those hard-scattering cross sections in QCD at large momentum transfer whose definitions are sufficiently inclusive.  This has opened  the way to many exact higher order and resumed cross sections.   Future progress will involve a bracing combination of precision calculations of inclusive cross sections, and an exploration of the role of long-distance properties, as we learn to relax the constraints imposed by inclusive cross sections, and explore the dynamics that opens up.

\end{document}